\documentclass[11pt,a4paper]{article}

\usepackage[utf8]{inputenc}
\usepackage[T1]{fontenc}
\usepackage{lmodern} 
\usepackage{amsmath,amssymb,amsthm,mathtools}
\usepackage{physics} 
\usepackage{microtype} 
\usepackage{geometry} 
\usepackage[dvipsnames]{xcolor} 
\usepackage{graphicx} 
\usepackage{cite} 
\usepackage{hyperref} 
\hypersetup{
    colorlinks=true,
    linkcolor=MidnightBlue,
    citecolor=MidnightBlue,
    urlcolor=RoyalBlue
}

\usepackage{booktabs}
\usepackage{array}
\usepackage{bm,bbm}
\usepackage{braket}
\usepackage{amsfonts}
\usepackage{dsfont}

\usepackage{authblk} 

\usepackage{titlesec}
\titleformat{\section}{\large\bfseries\color{black}}{\thesection}{1em}{}
\titleformat{\subsection}{\normalsize\bfseries\color{black}}{\thesubsection}{0.75em}{}

\theoremstyle{plain}

\theoremstyle{definition}

\theoremstyle{remark}

\title{\textbf{Symmetric orthogonalization and} \\ \textbf{probabilistic weights in resource quantification}}

\author[]{Gökhan Torun\thanks{email: \texttt{gokhan.torun@ieu.edu.tr}}}
\affil[]{\small Department of Physics, Faculty of Arts and Sciences, \\ Izmir University of Economics, 35330 Izmir, Turkiye}

\date{} 

\begin{document}

\maketitle

\begin{abstract}
Transforming non-orthogonal bases into orthogonal ones often compromises essential properties or physical meaning in quantum systems. Here, we demonstrate that Löwdin symmetric orthogonalization (LSO) outperforms the widely used Gram–Schmidt orthogonalization (GSO) in characterizing and quantifying quantum resources, with particular emphasis on coherence and superposition. We employ LSO both to construct an orthogonal basis from a non-orthogonal one and to obtain a non-orthogonal basis from an orthogonal set, thereby mitigating ambiguity related to the basis choice in defining quantum coherence. Unlike GSO, which depends on the ordering of input states, LSO applies a symmetric transformation that treats all vectors equally and minimizes deviation from the original basis. This procedure yields basis sets with enhanced stability, preserving the closest possible correspondence to the original physical states while satisfying orthogonality. Building on LSO, we also introduce Löwdin weights --- probabilistic weights for non-orthogonal representations that provide a consistent measure of resource content. We explicitly contrast these with Chirgwin-Coulson weights, demonstrating that Löwdin weights ensure non-negativity, a prerequisite for information-theoretic measures. These weights further enable the quantification of coherence and the characterization of superposition, providing a degree of superposition as a distinct measure, as well as facilitating the assessment of state delocalization through entropy and participation ratios. Our theoretical and numerical analyses confirm LSO’s superior preservation of quantum state symmetry and resource characteristics, underscoring the critical role of orthogonalization methods and Löwdin weights in resource theory frameworks involving non-orthogonal bases.
\end{abstract}



\section{Introduction}
\label{section-intro}

For decades, the problem of transforming linearly independent, non-orthogonal vectors into mutually orthogonal sets has been addressed within numerous theoretical and conceptual frameworks. Beyond simplifying computations, these transformations (i.e., orthogonalization) serve specialized roles in mathematics \cite{GramSchmidt100Years, Householder75}, physics \cite{Carlson57, Torun23Low}, and chemistry \cite{LOWDIN1950, LOWDIN1950SD}. Different methods excel in different scenarios: some prioritize computational efficiency and optimality \cite{GramSchmidt100Years, Carlson57, Zhang21}, while others emphasize exact physical constraints and geometric properties \cite{Torun23Low, LOWDIN1950, Srivastava00}. Consequently, the choice of orthogonalization method entails physical consequences, and understanding their nuances and comparative strengths remains an active and significant area of research.

Orthogonalization is more than a mathematical tool --- it possesses deep physical significance. In quantum systems, non-orthogonal basis states emerge naturally, as seen in atomic orbitals\footnote{Note that while atomic orbitals within a single atom are orthogonal, spatial overlap occurs between orbitals centered on different nuclei in molecular systems. In many-electron contexts, this leads to non-orthogonal Slater determinants or occupation vectors constructed from these overlapping orbitals.} \cite{LOWDIN1950}, squeezed states \cite{QSqueezing2003}, superposition states \cite{Theurer17}, and Schrödinger cat states \cite{CatState2007} (defined as superpositions of coherent states). Then, a critical consideration is how orthogonalizing such basis states affects their physical properties and those of the system. In (quantum) chemistry, orthogonalization plays a crucial role by enabling the construction of molecular orbitals from atomic bases. As discussed in Refs.~\cite{LOWDIN1950, LOWDIN1950SD}, enforced orthogonality ensures distinguishable electron states while preserving the local symmetry and geometric character of the original basis --- which is essential for an accurate physical interpretation of chemical bonding. Furthermore, orthogonalization reflects basic physical principles in quantum theory, such as the distinguishability of measurement outcomes. Importantly, the operational significance of orthogonalization is clearly demonstrated in the no-cloning theorem: arbitrary quantum states cannot be perfectly cloned \cite{NOCLO82} unless they belong to a set of mutually orthogonal states. In the realm of quantum resource theories \cite{Chitambar19, Regula19, Kuroiwa20, Torun23Review, Gour24}, the process of orthogonalization enables the quantification of coherence \cite{Baumgratz14} in a superposition of coherent states \cite{Tan17}. In another study, researchers showed that controlled orthogonalization --- enabled by quantum filtering --- becomes a powerful tool for manipulating quantum states while respecting the fundamental limits of quantum mechanics \cite{Jezek14}. Together, these perspectives show how orthogonalization operates as both a theoretical principle and a practical requirement across quantum theory.

Of the many useful orthogonalization techniques, the Gram-Schmidt orthogonalization (GSO) procedure \cite{GramSchmidt100Years} stands out as the most widely recognized and frequently employed method by researchers. A key advantage of this method lies in its applicability to countably infinite sets of vectors without requiring additional modifications. However, a notable limitation is that the resulting orthogonal set lacks a straightforward relationship to the original vectors. Specifically, the derived basis vectors depend on the sequential ordering of the input vectors, thereby introducing arbitrariness in certain applications.

While the Gram-Schmidt process remains commonly taught, Löwdin's two orthogonalization methods --- \emph{symmetric} and \emph{canonical} --- offer superior approaches for basis set transformation in (quantum) chemistry \cite{LOWDIN1950SD}. The symmetric method, \({\mathbf{E}} = {\mathbf{C}}\mathbf{O}_{d}^{-1/2}\) where \(\left[\mathbf{O}_{d}\right]_{ij} = \langle c_i | c_j \rangle\) is the overlap matrix of the non-orthogonal basis \({\mathbf{C}} = \left\{\ket{c_k}\right\}_{k=1}^{d}\), provides three key advantages \cite{LOWDIN1950, LOWDIN1950SD}: (1) it minimizes the orbital distortion \(\|{\mathbf{E}} - {\mathbf{C}}\|^2\) while enforcing exact orthonormality \(\langle e_i | e_j \rangle = \delta_{ij}\), (2) it preserves all molecular symmetries and degeneracies, and (3) it maintains maximal chemical interpretability of the resulting orbitals \({\mathbf{E}} = \left\{\ket{e_k}\right\}_{k=1}^{d}\). Unlike sequential orthogonalization methods, this symmetric approach treats all orbitals on an equal footing through a single transformation which is linear and invertible. The canonical variant, while useful for ill-conditioned systems (the smallest eigenvalue \(\lambda_{\min}\) of the overlap matrix satisfies \(\lambda_{\min} \to 0\)) via eigendecomposition of \(\mathbf{O}_{d}\), often compromises orbital symmetry. Consequently, Löwdin symmetric orthogonalization (LSO) has become indispensable in wavefunction-based methods, where its mathematical elegance (derived from the optimality of \(\mathbf{O}_{d}^{-1/2}\)) and physical relevance converge to yield chemically meaningful orbitals. Notably, a key feature of LSO is its ability to transform maximally coherent states \cite{Baumgratz14} into maximal superposition states \cite{Torun21, Senyasa22} while preserving their essential quantum properties, as established in Ref.~\cite{Torun23Low}. This unique capability arises from LSO's mathematical structure, which establishes a direct link between these quantum resources \cite{Baumgratz14, Theurer17}. Therefore, the LSO method is expected to provide significant utility for addressing quantum resource quantification and characterization problems.

In this paper, we address the problem of quantifying and characterizing quantum resources by means of the LSO approach, taking advantage of its inherent benefits, with a particular focus on coherence and superposition. As is well known, quantum coherence \cite{Baumgratz14} is defined relative to a fixed orthonormal basis, whereas superposition \cite{Theurer17} is characterized with respect to non-orthogonal basis sets. This distinction poses challenges in consistently analyzing and comparing these quantum resources within a unified framework. For example, while there exists a considerable number of resource measures for coherence \cite{Streltsov2017QuCo}, extending these measures to superposition is not straightforward. A key contribution of this study lies in the application of LSO to address this challenge, thereby partially mitigating it and offering guidance for subsequent investigations. Specifically, we demonstrate that standard coherence measures can be applied to the Löwdin-transformed state to explicitly quantify the coherence content of a superposition state. This methodology facilitates a simple characterization of the given superposition state, circumventing the ambiguities typically associated with non-orthogonal frameworks. In doing so, we establish a formal definition for the degree of superposition relative to the original basis. Furthermore, our objective includes promoting broader awareness of Löwdin orthogonalization (LO) to complement the established GSO approach.

The paper is organized as follows. In Section \ref{section-Techniques}, we review relevant orthogonalization methods. First, we discuss the Gram–Schmidt method (Section \ref{section-GramSchmidt}), highlighting its ubiquity and inherent limitations. Subsequently, we detail Löwdin’s orthogonalization techniques (Section \ref{section-Löwdin}), with a focus on the symmetric (Section \ref{section-LSymmetric}) and canonical (Section \ref{section-LCanonical}) approaches. Although these methods are closely related, they differ fundamentally in their suitability for quantum resource analysis: LSO preserves the symmetry of the original basis, whereas canonical orthogonalization diagonalizes the overlap matrix to yield an orthonormal basis aligned with its eigenvectors. We also demonstrate how LSO can be used to construct a non-orthogonal basis from an orthogonal one, such as the computational basis (Section \ref{section-ComptBasis}). This bidirectional construction via LSO resolves ambiguities regarding basis selection for defining coherence or superposition. In Section \ref{section-QResources}, we substantiate this approach by showing that the physical description of the system relies primarily on the overlap matrix and superposition coefficients, rather than the explicit spatial forms of the non-orthogonal basis vectors. Building on this premise, Section \ref{section-LÖwdinWeights} formalizes the concept of Löwdin weights for both pure and mixed states. These weights provide consistent probability assignments for non-orthogonal representations, offering a robust framework for evaluating quantum resources. In Section \ref{section-LÖwdinWeights-Examples}, we apply this framework to illustrative examples to rigorously characterize superposition states, quantifying the coherence content and establishing a distinct measure for the degree of superposition. In Section \ref{section-EntropicandGeometric}, we extend this analysis to the entropic and geometric quantification of localization, offering a complementary physical interpretation of the Löwdin weights. Our theoretical and numerical results establish that LSO preserves the symmetry and key features of quantum states more effectively than GSO. Finally, in Section \ref{section-Conc}, we summarize our main findings and suggest directions for future work. We emphasize that our results provide critical insights into the role of orthogonalization in quantum resource theory \cite{Chitambar19, Regula19, Kuroiwa20, Torun23Review, Gour24} and facilitate the development of advanced methodologies to analyze coherence and superposition \cite{Baumgratz14, Theurer17, Streltsov2017QuCo}.


\section{Techniques for Orthogonalization}
\label{section-Techniques}

In this section, we examine the two orthogonalization methods mentioned earlier. We begin with the Gram-Schmidt process. Next, we present Löwdin’s symmetric method, followed by its canonical variant. After that, we contextualize Löwdin’s symmetric technique within the computational basis to facilitate the later comparisons of quantum resources (e.g., coherence \cite{Baumgratz14} and superposition \cite{Theurer17}). A brief summary of these techniques is provided in Table~\ref{table1}, highlighting their relative strengths and limitations.

\begin{table}[htb!]
    \centering
    \caption{Comparative analysis of Gram-Schmidt and LSO schemes. The Löwdin method is distinguished by its global, order-independent nature and the minimization of the least-squares distance between the orthogonal set $\{|e_i\rangle\}$ and the original basis $\{|c_i\rangle\}$.}
    \label{table1}
    \renewcommand{\arraystretch}{1.3} 
    \setlength{\tabcolsep}{8pt}      
    \begin{tabular}{@{} l l l @{}}    
    \toprule
    \textbf{Property} & \textbf{Gram-Schmidt} & \textbf{Löwdin Symmetric} \\ 
    \midrule
    Transformation & Sequential; Non-unitary & Global; Hermitian; Invertible; Linear \\ 
    Numerical Stability & Moderate & High (if overlap matrix is well-conditioned) \\ 
    Order Dependence & Yes & No (basis-invariant) \\ 
    Minimal Distortion & No & Yes (minimizes $\sum_i \|e_i - c_i\|^2$) \\ 
    Coherence Control & Poor (Asymmetric) & Optimal (Symmetric) \\ 
    \bottomrule
    \end{tabular}
\end{table}


\subsection{Gram-Schmidt orthogonalization}
\label{section-GramSchmidt}

As one of the most well-known orthogonalization techniques, the Gram-Schmidt process is a staple of linear algebra due to its straightforward iterative structure \cite{GramSchmidt100Years}. Owing to its broad familiarity, historical significance, and extensive applications in mathematics, physics, and engineering, it serves as a natural reference point in discussions of basis transformations. For these reasons, we give it brief consideration here and compare it with the LSO, as shown in Table~\ref{table1}.

Let \(\left\{\ket{c_{k}}\right\}_{k=1}^{d}\) be a linearly independent set of non-orthogonal vectors in an inner product space. The orthonormal set \(\left\{\ket{{e}_{k}^{\text{GS}}}\right\}_{k=1}^{d}\) with the same span is constructed via the Gram-Schmidt process, henceforth referred to as the Gram-Schmidt basis, as follows:
\begin{align}
    \mathbf{v}_1 &= \ket{c_{1}}, \quad \ket{{e}_{1}^{\text{GS}}} = \frac{\mathbf{v}_1}{\|\mathbf{v}_1\|}; \label{GramOrth1} \\
    \mathbf{v}_2 &= \ket{c_{2}} - \langle{c_{2}} | {{e}_{1}^{\text{GS}}}\rangle  \ket{{e}_{1}^{\text{GS}}}, \quad \ket{{e}_{2}^{\text{GS}}} = \frac{\mathbf{v}_2}{\|\mathbf{v}_2\|}; \label{GramOrth2} \\
    \mathbf{v}_j &= \ket{c_{j}} - \sum_{i=1}^{j-1} \langle{c_{j}} | {{e}_{i}^{\text{GS}}}\rangle  \ket{{e}_{i}^{\text{GS}}}, \quad \ket{{e}_{j}^{\text{GS}}} = \frac{\mathbf{v}_j}{\|\mathbf{v}_j\|} \quad \text{for } j = 3,\ldots,d. \label{GramOrthd}
\end{align}
Here \(\mathbf{v}_j\) in Eq.~\eqref{GramOrthd} is the orthogonalized (but unnormalized, in general) vector at step \(j\), and \(\|\mathbf{\cdot}\|\) is the inner product norm. The GSO process ensures orthogonality by subtracting, at each step, the components of \(\ket{{c}_{j}}\) along the prior orthonormal vectors \(\left\{\ket{{e}_{i}^{\text{GS}}}\right\}_{i=1}^{j-1}\), ensuring orthogonality. Once normalized, the orthogonalized \(\mathbf{v}_k\) form unit vectors, completing an orthonormal basis \(\left\{\ket{{e}_{k}^{\text{GS}}}\right\}_{k=1}^{d}\).

\begin{figure}[h!]
\centering
\includegraphics[width=13.00cm]{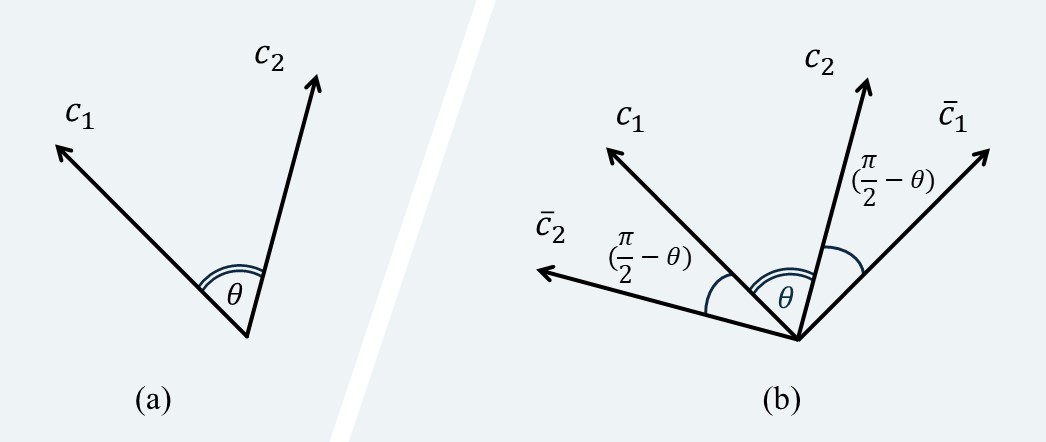}
\caption{Order dependence of the GSO process. (a) Two non-orthogonal vectors \(\left\{\ket{c_1} (=c_1), \ket{c_2} (=c_2)\right\}\) with relative angle \(\theta \in (0, \pi/2)\). (b) Orthogonalization results differ based on which vector is processed first: Starting with \(\ket{c_1}\) yields \(\left\{\ket{{e}_{1}^{\text{GS}}} (={c_1}), \ket{{e}_{2}^{\text{GS}}} (={\bar{c}_1})\right\}\); Starting with \(\ket{c_2}\) gives \(\left\{\ket{{e}_{1}^{\text{GS}}} (={c_2}), \ket{{e}_{2}^{\text{GS}}} (={\bar{c}_2})\right\}\). This illustrates the intrinsic order dependence of Gram-Schmidt, the two bases are distinct. The same applies in higher dimensions. For clarity, all vectors are assumed unit-norm.}
\label{FIG:GramSchmidt}
\end{figure}

On the one hand, the GSO approach offers algorithmic simplicity, allowing for efficient sequential vector processing. On the other hand, its most critical drawback is arguably order dependence, since the resulting orthogonal basis is highly sensitive to the input vector sequence (see Figure~\ref{FIG:GramSchmidt}). This introduces undesirable arbitrariness in applications where basis symmetry is physically meaningful, compromising the reliability of the orthogonalized representation. Consequently, there is indeed a well-motivated need to explore alternative orthogonalization methods that mitigate this limitation while preserving computational tractability.

\subsection{L\"{o}wdin's orthogonalization methods}
\label{section-Löwdin}

Although less widely known than GSO in general, Löwdin's orthogonalization method proves especially powerful in applications where symmetry preservation is critical \cite{LOWDIN1950, LOWDIN1950SD, Torun23Low}. Let us return to the origin of the problem and consider the given set of  linearly independent but non-orthogonal basis vectors \(\left\{\ket{c_{k}}\right\}_{k=1}^{d}\) in a Hilbert space. The non-orthogonality is characterized by the overlap matrix \(\mathbf{O}_{d}\) (also called the Gram matrix):
\begin{equation}\label{OverlapMat}
\mathbf{O}_{d} = 
\begin{pmatrix}
\langle c_1 | c_1 \rangle & \langle c_1 | c_2 \rangle & \cdots & \langle c_1 | c_d \rangle \\
\langle c_2 | c_1 \rangle & \langle c_2 | c_2 \rangle & \cdots & \langle c_2 | c_d \rangle \\
\vdots & \vdots & \ddots & \vdots \\
\langle c_d | c_1 \rangle & \langle c_d | c_2 \rangle & \cdots & \langle c_d | c_d \rangle
\end{pmatrix},
\end{equation}
where \(O_{ij} (\equiv \left[\mathbf{O}_{d}\right]_{ij}) = \langle c_i | c_j \rangle\) explicitly represents the \((i,j)\)-th element of matrix \(\mathbf{O}_{d}\). The overlap matrix \(\mathbf{O}_{d}\) is Hermitian (\(\mathbf{O}_{d}^\dagger = \mathbf{O}_{d}\)) and positive-definite. It should be emphasized that, when employing non-orthogonal basis sets, the overlap matrix plays a crucial role in preserving the correct geometric and physical structure of the system. Through the use of the overlap matrix, we construct the orthonormal basis states \(\left\{\ket{{e}_{k}^{\text{L}}}\right\}_{k=1}^{d}\), commonly referred to as the Löwdin basis.

Now consider the case where the sets \(\left\{\ket{c_{k}}\right\}_{k=1}^{d}\) and \(\left\{\ket{{e}_{k}^{\text{L}}}\right\}_{k=1}^{d}\) span the same subspace of the Hilbert space. In this case, the orthogonal states \(\ket{{e}_{i}^{\text{L}}}\) can be expressed as linear combinations of the original non-orthogonal states \(\left\{\ket{c_{k}}\right\}_{k=1}^{d}\) through the (linear) transformation
\begin{equation}
\ket{{e}_{i}^{\text{L}}} = \sum_{k=1}^d T_{ki} \ket{c_{k}}, \quad i = 1, \ldots, d,
\end{equation}
where the coefficients \(T_{ki} \equiv \left[\mathbf{T}\right]_{ki}\) form a \(d \times d\) transformation matrix \(\mathbf{T}\). The problem is then formulated as determining the transformation coefficients \(\left\{T_{ki}\right\}\) via the overlap matrix \(\mathbf{O}_{d}\) given by Eq.~\eqref{OverlapMat}. Per-Olov Löwdin introduced two closely related but distinct approaches \cite{LOWDIN1950, LOWDIN1950SD} for constructing orthonormal bases from non-orthogonal ones, namely, symmetric orthogonalization (SO) and canonical orthogonalization (CO).


\subsubsection{Symmetric orthogonalization}
\label{section-LSymmetric}

This method produces an orthonormal basis \(\left\{\ket{{e}_{i}^{\text{L}}}_{\mathrm{sym}}\right\}_{i=1}^{d}\) that is the ``closest'' to the original basis in a least-squares sense. One achieves this through the following three steps:
\begin{enumerate}
\item \emph{Diagonalize the overlap matrix}: Solve the eigenvalue problem \(\mathbf{O}_{d} \mathbf{u}_k = \lambda_k\mathbf{u}_k\) for \(k=1,\ldots,d\), where \(\lambda_k > 0\) (since \(\mathbf{O}_{d}\) is positive definite) and \(\{\mathbf{u}_k\}\) form an orthonormal set.

\item \emph{Construct the inverse square root}: Form the eigenvector matrix \(\mathbf{U} = (\mathbf{u}_1 \ldots \mathbf{u}_d)\) and diagonal eigenvalue matrix \(\mathbf{D} = \text{diag}(\lambda_1, \ldots, \lambda_d)\). Then compute
\begin{equation}
\mathbf{T}_{\mathrm{sym}} = \mathbf{O}_{d}^{-1/2} = \mathbf{U}\mathbf{D}^{-1/2}\mathbf{U}^\dagger \quad \text{where} \quad \mathbf{D}^{-1/2} = \text{diag}\left(\frac{1}{\sqrt{\lambda_1}}, \ldots, \frac{1}{\sqrt{\lambda_d}}\right).
\end{equation}

\item \emph{Apply the transformation}: The (symmetric) orthonormal basis vectors are then given by
\begin{equation}\label{Eq:LöwdinSymmDef}
\ket{{e}_{i}^{\text{L}}} = \ket{{e}_{i}^{\text{L}}}_{\mathrm{sym}} = \sum_{k=1}^{d} \left[\mathbf{O}_{d}^{-1/2}\right]_{ki} \ket{c_k}, \quad i = 1, \ldots, d.
\end{equation}
\end{enumerate}

\begin{figure}[h!]
\centering
\includegraphics[width=13.00cm]{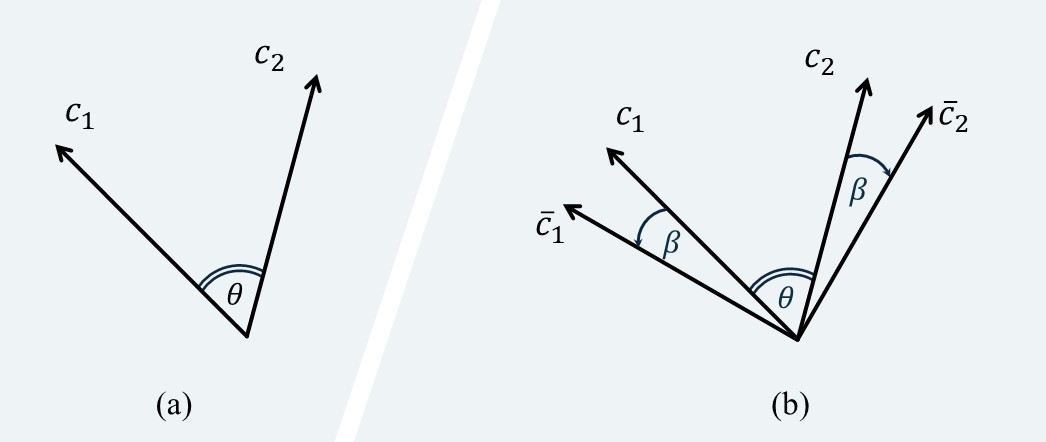}
\caption{Geometric illustration of Löwdin SO for a two-dimensional system. (a) The original non-orthogonal vectors \(\ket{c_1} (=c_1)\) and \(\ket{c_2} (=c_2)\) exhibit a nonzero overlap (with \(\theta \in (0, \pi/2)\)). (b) After symmetrically rotating both vectors by an angle \(\beta\), they are transformed into the orthonormal Löwdin basis \(\ket{{e}_{1}^{\text{L}}}_{\mathrm{sym}} (=\bar{c}_1)\) and \(\ket{{e}_{2}^{\text{L}}}_{\mathrm{sym}} (=\bar{c}_2)\), where \(\theta +2\beta = \pi/2\) ensures mutual orthogonality. The same applies in higher dimensions. All vectors are assumed unit-norm.}
\label{FIG:LöwdinSymmetric}
\end{figure}

The (Löwdin) SO process described by Eq.~\eqref{Eq:LöwdinSymmDef} stands out as the preferred basis transformation because it uniquely meets three essential criteria. First, it exactly minimizes the Frobenius norm \cite{LOWDIN1950, LOWDIN1950SD}
\begin{equation}\label{Least-Square-Sense}
\|{\mathbf{E}} - {\mathbf{C}}\|_{\mathrm{F}} = \sqrt{\sum_k \|{e}_{k}^{\text{L}} - c_k\|^2},
\end{equation}
ensuring the orthogonalized basis \({\mathbf{E}} = \left\{\ket{{e}_{k}^{\text{L}}}_{\mathrm{sym}}\right\}_{k=1}^{d}\) preserves maximal geometric similarity to the initial non-orthogonal basis \({\mathbf{C}} = \left\{\ket{c_{k}}\right\}_{k=1}^{d}\) (see Figure~\ref{FIG:LöwdinSymmetric}). Second, through the transformation \eqref{Eq:LöwdinSymmDef}, SO conserves all inherent symmetries of \({\mathbf{C}}\) (a critical feature for quantum chemistry and condensed matter applications where symmetry-adapted bases are essential). Third, in contrast to sequential methods, which are sensitive to the choice of the first vector (e.g., Gram-Schmidt), SO preserves the equal status of basis states via the symmetric form of \(\mathbf{O}_{d}^{-1/2}\), which applies a uniform transformation to all vectors in \({\mathbf{C}}\). We might conclude that this triad of properties --- minimal distortion, symmetry preservation, and basis democracy --- establishes SO as one of the optimal orthogonalization processes for quantum mechanical bases. In molecular systems, for instance, Löwdin symmetrically-orthogonalized orbitals retain maximal correspondence with atomic orbitals while satisfying orthogonality constraints \cite{LOWDIN1950, LOWDIN1950SD}. However, it is worth noting that while Löwdin symmetrically-orthogonalized orbitals retain maximal correspondence with atomic orbitals, they do inevitably introduce some spatial delocalization (``tails'') across neighboring centers to satisfy the orthogonality constraints.

\subsubsection{Canonical orthogonalization}
\label{section-LCanonical}

This method produces an orthonormal basis \(\left\{\ket{{e}_{i}^{\text{L}}}_{\mathrm{can}}\right\}_{i=1}^{d}\) and it is particularly useful when some eigenvalues \(\lambda_k\) are small (indicating near-linear dependence, i.e., ill-conditioned systems). One achieves this through the following three steps:

\begin{enumerate}
\item \emph{Diagonalize the overlap matrix}: Obtain the eigenvector matrix \(\mathbf{U} = (\mathbf{u}_1 \ldots \mathbf{u}_d)\) and diagonal eigenvalue matrix \(\mathbf{D} = \text{diag}(\lambda_1, \ldots, \lambda_d)\) from the eigenvalue problem, \(\mathbf{O}_{d} \mathbf{u}_k = \lambda_k\mathbf{u}_k\).

\item \emph{Form the transformation matrix}: Construct \(\mathbf{T}_{\mathrm{can}} = \mathbf{U}\mathbf{D}^{-1/2}\). Notice that this differs from SO by simply omitting \(\mathbf{U}^\dagger\).

\item \emph{Generate the new basis}: The (canonical) orthonormal vectors are then given by 
\begin{equation}\label{Eq:LöwdinCanDef}
\ket{{e}_{i}^{\text{L}}} = \ket{{e}_{i}^{\text{L}}}_{\mathrm{can}} = \sum_{k=1}^d \frac{U_{ki}}{\sqrt{\lambda_k}} \ket{c_k}, \quad i=1,\ldots,d.
\end{equation}
\end{enumerate}

The (Löwdin) CO process described by Eq.~\eqref{Eq:LöwdinCanDef} directly aligns with the eigenvectors of \(\mathbf{O}_{d}\), making it preferable for systems with near-linear dependence (i.e., small eigenvalues \(\lambda_k\)). As it seen, SO requires the full matrix product \(\mathbf{T}_{\mathrm{sym}} = \mathbf{U}\mathbf{D}^{-1/2}\mathbf{U}^\dagger\), whereas CO only needs the transformation \(\mathbf{T}_{\mathrm{can}} = \mathbf{U}\mathbf{D}^{-1/2}\). These yield identical results when \(\lambda_k \approx 1\), but diverge when \(\mathbf{O}_{d}\) contains small eigenvalues or strong non-orthogonalities. In summary, the symmetric approach \eqref{Eq:LöwdinSymmDef} preserves the original basis relationships, making it ideal for quantum mechanical applications where physical interpretability is important. In contrast, the canonical method \eqref{Eq:LöwdinCanDef} offers inherent stability with near-singular overlap matrices, making it better suited for numerical applications that require robust eigenvector alignment. Accordingly, LSO serves as a key tool in our calculations.


\subsection{Fixing the Löwdin basis as the computational basis}
\label{section-ComptBasis}

As demonstrated, the SO process transforms a non-orthogonal basis \({\mathbf{C}}\) into an orthogonal basis \({\mathbf{E}}\) via the transformation given in Eq.~\eqref{Eq:LöwdinSymmDef}. Alternatively, the transformation \eqref{Eq:LöwdinSymmDef} can also be expressed as
\begin{equation}\label{Eq:PhitoPsi}
\mathbf{E} = \mathbf{C} \mathbf{O}_d^{-1/2}.
\end{equation}
Since \(\mathbf{O}_d\) is Hermitian, \(\mathbf{O}_d^{-1/2}\) is also Hermitian. The transformation \eqref{Eq:PhitoPsi} guarantees orthonormality of \(\mathbf{E}\). Note that the matrix \(\mathbf{O}_d^{-1/2}\) is invertible due to the positive definiteness of \(\mathbf{O}_d\) (\(\lambda_k > 0\)). The inverse square root of \(\mathbf{O}_d\) multiplied by its square root yields the identity matrix, \(\mathbf{O}_d^{-1/2} \mathbf{O}_d^{1/2} = \mathbf{I}_{d}\). The inverse transformation is therefore well-defined:
\begin{equation}\label{Eq:PsitoPhi}
\mathbf{C} = \mathbf{E} \mathbf{O}_d^{1/2}.
\end{equation}
This framework is valid for any finite-dimensional basis in physical quantum systems, where \(\mathbf{O}_d\) is always positive definite with strictly positive eigenvalues.

Expressing the relation \eqref{Eq:PsitoPhi} in terms of individual basis vectors, we obtain
\begin{equation}\label{Eq:LöwdinSymmDef2}
\ket{c_i} = \sum_{k=1}^{d} \left[\mathbf{O}_{d}^{1/2}\right]_{ki} \ket{e_{k}^{\text{L}}}_{\mathrm{sym}}, \quad i = 1, \ldots, d,
\end{equation}
where \(\left\{\ket{e_{k}^{\text{L}}}_{\mathrm{sym}}\right\}_{k=1}^{d}\) denotes the L\"owdin (symmetrically orthogonalized) basis. For computational convenience, we may represent the orthogonalized basis \({\mathbf{E}}\) in terms of a standard computational basis:
\begin{equation}\label{Eq:Computational}
\left\{\ket{e_{k}^{\text{L}}}_{\mathrm{sym}}\right\}_{k=1}^{d} = \left\{\ket{1}, \ket{2}, \ldots, \ket{d}\right\},
\end{equation}
where \(\{\ket{k}\}_{k=1}^{d}\) are orthonormal states spanning the \(d\)-dimensional Hilbert space. This choice, without loss of generality, not only simplifies computations but also facilitates the study of quantum resources across different basis representations. Specifically, by treating the computational basis as the original orthonormal set and applying LSO inversely to generate symmetrically non-orthogonalized vectors (Eq.~\eqref{Eq:PsitoPhi}), one can probe superposition more deeply, extending coherence theory to a broader framework. In the following sections, we bring together all the information presented above to derive novel results beyond what has been previously established for coherence \cite{Baumgratz14} and superposition \cite{Theurer17}.


\section{Quantum Resources in Relation to Reference Bases}
\label{section-QResources}

Quantum resource theories (QRTs) have come to be a study of how quantum properties could lead to an operational advantage in quantum information processing, and broadly speaking, are concerned with their quantification, characterization, and manipulation under physical restrictions \cite{Chitambar19, Regula19, Kuroiwa20, Torun23Review, Gour24}. A resource theory is composed of two elements: the free states (whose set is denoted by \({\mathcal{F}}\)) and the free operations (\({\mathcal{O}}\)). All states which are not free are regarded as (quantum) resources. The free operations are physical transformations which do not create any resources; that is, they must transform free states into free ones, allowing for the resource to be manipulated but not freely created. For an extended discussion of these themes, see Refs.~\cite{Chitambar19, Regula19, Kuroiwa20, Torun23Review, Gour24}.

The primary focus of our work is the \emph{superposition} of linearly independent vectors. Depending on whether the preferred basis vectors are orthogonal or non-orthogonal, different QRTs can be formulated. In this context, two closely related QRTs have been developed: the resource theory of coherence (RTC) \cite{Baumgratz14}, defined with respect to an orthogonal basis, and the resource theory of superposition (RTS) \cite{Theurer17}, which generalizes the former to non-orthogonal basis sets. While both theories quantify quantum resources arising from superpositions, the non-orthogonal case requires additional considerations due to the nonvanishing overlaps between basis states.

Quantum coherence \cite{Baumgratz14} is inherently a basis-dependent property. Whether a quantum state exhibits coherence depends on the choice of the orthonormal basis with respect to which coherence is defined. To illustrate this, consider two orthonormal basis sets in a two-dimensional Hilbert space: the computational basis \(\left\{\ket{0}, \ket{1}\right\}\) and the Hadamard basis \(\left\{\ket{+}, \ket{-}\right\}\), where \(\ket{\pm} = \left(\ket{0} \pm \ket{1}\right)/{\sqrt{2}}\).
Then, the coherence property of the state \(\ket{\chi} = \left(\ket{0} + \ket{1}\right)/{\sqrt{2}} \: (\equiv \ket{+})\) depends on the basis chosen. Obviously, the state \(\ket{\chi}\) exhibits coherence in the computational basis but is incoherent in the Hadamard basis. Specifically, a state that is a superposition in one basis may be an eigenstate (and hence incoherent) in another. Consequently, any rigorous discussion of quantum coherence requires the explicit specification of the reference basis. Following our discussion of the computational basis (equivalently, the Löwdin basis) in Section \ref{section-ComptBasis}, we formally adopt this as the fixed reference basis for all subsequent coherence and superposition analysis.

The RTS \cite{Theurer17} employs a set of non-orthogonal yet linearly independent basis states \(\left\{\ket{c_{k}}\right\}_{k=1}^{d}\) for a \(d\)-dimensional Hilbert space \(\mathcal{H}_d\). Within this framework, superposition-free states are precisely those of the form $\rho = \sum_k \rho_k \ket{c_k}\bra{c_k}$, where $\rho_k \geq 0$ and $\sum_k \rho_k = 1$. The set of resource states (those exhibiting superposition) consists of all density operators outside \(\mathcal{F}\). Pure superposition states are characterized by linear combinations of the form \(a_1 \ket{c_1} + a_2 \ket{c_2} + \dots + a_n \ket{c_{n}}\) (with \(a_i \neq 0\) for \(i=1, 2, \dots, n \in (2, d]\)), subject to the normalization condition \(\sum_{i, j=1}^{d} a_{i}^{\ast} O_{ij} a_{j}=1\). Despite their apparent similarity, coherence and superposition exhibit an obvious distinction --- the former is formulated in an orthogonal basis, while the latter is formulated in a non-orthogonal basis. As mentioned, this implies that superposition theory constitutes a generalization of coherence theory \cite{Theurer17}.

As established in \cite{Torun23Low}, the Löwdin SO method serves as an effective tool for characterizing pure superposition states. As discussed in Section \ref{section-LSymmetric}, a defining feature of SO is its ability to maximally preserve the structure and symmetry of the original non-orthogonal basis states. This property leads to an investigation of Löwdin SO's role in determining the hierarchical relationships among resource states. Crucially, as shown in \cite{Torun23Low}, states of maximal superposition \cite{Senyasa22} are mapped to maximally coherent states under SO, and vice versa --- a consequence of the invertible nature of LSO. Thus, maximally coherent states and states with maximal superposition are identified as equivalent under the action of Löwdin's SO. To make this important and interesting result more concrete, let us examine the case where \(d=2\). Following the steps from Section \ref{section-LSymmetric}, the Löwdin basis states can be obtained in the following form:
\begin{eqnarray}\label{E1-Löwdin}
\ket{e_{1}^{\text{L}}}_{\mathrm{sym}} \equiv \ket{1} = \frac{1}{2} \left(\frac{1}{\sqrt{\lambda_2}} + \frac{1}{\sqrt{\lambda_1}}\right) \ket{c_1} + \frac{1}{2} \left(\frac{1}{\sqrt{\lambda_2}} - \frac{1}{\sqrt{\lambda_1}}\right) \ket{c_2},
\end{eqnarray}
\begin{eqnarray}\label{E2-Löwdin}
\ket{e_{2}^{\text{L}}}_{\mathrm{sym}} \equiv \ket{2} = \frac{1}{2} \left(\frac{1}{\sqrt{\lambda_2}} - \frac{1}{\sqrt{\lambda_1}}\right) \ket{c_1} + \frac{1}{2} \left(\frac{1}{\sqrt{\lambda_2}} + \frac{1}{\sqrt{\lambda_1}}\right) \ket{c_2},
\end{eqnarray}
where \(\lambda_2 = 1+s\), \(\lambda_1 = 1-s\), and \(\langle c_1 | c_2 \rangle = s \in (-1, 1)\). Additionally, based on the derivations provided in Section \ref{section-ComptBasis}, the relations
\begin{eqnarray}\label{C1-Computational}
\ket{c_1} = \frac{1}{2} \left[\left({\sqrt{\lambda_2}} + {\sqrt{\lambda_1}}\right) \ket{1} + \left({\sqrt{\lambda_2}} - {\sqrt{\lambda_1}}\right) \ket{2} \right]
\end{eqnarray}
and
\begin{eqnarray}\label{C2-Computational}
\ket{c_2} = \frac{1}{2} \left[\left({\sqrt{\lambda_2}} - {\sqrt{\lambda_1}}\right) \ket{1} + \left({\sqrt{\lambda_2}} + {\sqrt{\lambda_1}}\right) \ket{2} \right]
\end{eqnarray}
are immediately derived. Consider the maximally coherent states in a two-dimensional system, given by \(\left(\ket{1} \pm \ket{2}\right)/\sqrt{2}\). Substituting Eqs.~\eqref{E1-Löwdin} and \eqref{E2-Löwdin} into the maximally coherent states leads to:
\begin{eqnarray}
\frac{1}{\sqrt{2}} \big(\ket{1} + \ket{2}\big)  \longrightarrow \frac{1}{\sqrt{2\lambda_2}}\big(\ket{c_1} + \ket{c_2}\big) \quad \text{where} \quad  \langle c_1 | c_2 \rangle = s \in (-1, 0]
\end{eqnarray}
and
\begin{eqnarray}
\frac{1}{\sqrt{2}} \big(\ket{1} - \ket{2}\big) \longrightarrow \frac{1}{\sqrt{2\lambda_1}}\big(\ket{c_1} - \ket{c_2}\big) \quad \text{where} \quad  \langle c_1 | c_2 \rangle = s \in [0, 1).
\end{eqnarray}
This result carries substantial importance, given that overlap effects introduce complexity in analyzing superposition states. Accordingly, the Löwdin SO process stands out as a valuable analytical tool. We conclude this discussion here; further details are available in \cite{Torun23Low}.

We now turn our attention to another important question: In a given superposition state, does the explicit form of the non-orthogonal basis states hold fundamental significance, or is their overlap the decisive factor? To address this, we analyze a two-dimensional system, comparing three distinct sets of non-orthogonal bases. As an illustrative example, consider the following normalized superposition state:
\begin{eqnarray}\label{StatePhiBasisIndep}
\ket{\varphi} = \ket{c_1} - \left(\frac{2+\sqrt{2}}{\sqrt{3}}\right)\ket{c_2}.
\end{eqnarray}
Here, the overlap between basis states \(\ket{c_1}\) and \(\ket{c_2}\) is \(\left(1+\sqrt{2}\right)/\sqrt{6}\). This basis set admits multiple equivalent representations. For instance,
\begin{eqnarray}
  S^{(1)}&:& \left\{\ket{c_1} = \frac{1}{\sqrt{2}}\big(\ket{1} + \ket{2}\big), \ket{c_2} = \sqrt{\frac{2}{3}}\ket{1} + \frac{1}{\sqrt{3}}\ket{2}\right\},  \\
  S^{(2)}&:& \left\{\ket{c_1} = \ket{1}, \ket{c_2} = \frac{1+\sqrt{2}}{\sqrt{6}}\ket{1} + \frac{1-\sqrt{2}}{\sqrt{6}}\ket{2}\right\}, \\
  S^{(3)}&:& \left\{\text{Eq.~\eqref{C1-Computational}},  \text{Eq.~\eqref{C2-Computational}} \right\}.
\end{eqnarray}
For each basis set \( S^{(i)}\) (\(i=1, 2, 3\)), the overlap \(\langle c_1 | c_2 \rangle = s\) is fixed at \(\left(1+\sqrt{2}\right)/\sqrt{6}\). This consistency demonstrates that physical predictions depend solely on the overlap, not the explicit basis representation. Consequently, integrating the Löwdin SO poses no theoretical barrier; in fact, it offers practical advantages by preserving the original overlap structure while simplifying calculations. Crucially, this insight underscores a broader principle: fixation on explicit representations, when overlaps are known, can introduce unnecessary computational overhead. For instance, basis-specific parameterizations may require additional transformations or introduce numerical instability, whereas overlap-driven methods (like Löwdin’s) bypass these pitfalls. To systematically test this, we adopt computational reference bases (as the Löwdin basis) and generate non-orthogonal sets with controlled overlaps (this is implemented via Eq.~\eqref{Eq:PsitoPhi}). This approach isolates the role of overlaps in superposition analyses, avoiding artifacts from arbitrary coordinate choices.

Having clarified this point, we now address a deeper and equally critical question: What is the degree of coherence in a given superposition state? This inquiry lies at the heart of our study, as coherence quantifies quantum behavior in such systems where non-orthogonal basis states are constructed from the Löwdin basis (specifically, the computational basis in our case). In the following sections, we resolve this question by integrating analytical reasoning with explicit examples.


\section{Löwdin Weights}
\label{section-LÖwdinWeights}

Measurement probabilities of a quantum state are determined by the modulus squared of its expansion coefficients in an orthonormal basis \cite{Brumer2006, Zurek2011}. However, in the presence of non-orthogonal bases, such as atomic orbitals in quantum chemistry or coherent states in quantum optics, a more careful treatment is required. Importantly, the coefficients in such expansions are not directly interpretable as probabilities due to interference effects and overlap between basis states. We explore this problem through the use of Löwdin's SO process. This construction defines a new class of quantities, namely the Löwdin weights, which form a probability distribution over the symmetrically orthogonalized basis \cite{LowdinW1953}.

Consider a normalized (pure) quantum superposition state \(\ket{\alpha} \in \mathrm{span}\left\{\ket{c_k}\right\}_{k=1}^{d}\), expressed as
\begin{eqnarray}\label{Sup-Alpha}
\ket{\alpha}= \sum_{k=1}^{d} a_k \ket{c_k}, \quad a_k \in \mathds{C}.
\end{eqnarray}
The LSO process constructs an orthonormal basis \(\left\{\ket{e_{k}^{\text{L}}}_{\mathrm{sym}}\right\}_{k=1}^{d}\) from the original non-orthogonal set by applying Eq.~\eqref{Eq:LöwdinSymmDef}. To express \(\ket{\alpha}\) in this orthonormal basis, we define the coefficient vector \(\mathbf{a} = \left(a_1, \dots, a_d\right)^{\mathrm{T}} \in \mathds{C}^{d}\). The representation of \eqref{Sup-Alpha} in the Löwdin basis (i.e., computational basis) is then given by
\begin{eqnarray}\label{LöwdinWeightsDefinition}
\ket{\alpha} = \sum_{k=1}^{d} b_k \ket{e_{k}^{\text{L}}}_{\mathrm{sym}} \equiv \sum_{k=1}^{d} b_k \ket{k}, \quad \text{where} \quad \mathbf{b} = \left(b_1, \dots, b_d\right)^{\mathrm{T}} = \mathbf{O}_{d}^{1/2} \mathbf{a}.
\end{eqnarray}
Once the vector \(\mathbf{b}\) is obtained (physically, the coefficients \(\left\{b_k\right\}_{k=1}^{d}\) in the Löwdin basis can be seen as renormalized contributions of the original non-orthogonal basis states), the weight contribution of the \(k\)-th orthogonalized basis vector to the state \(\ket{\alpha}\) is computed as follows:
\begin{equation}\label{LöwdinWeightsD}
w_k^{\text{L}}(\alpha) = |b_k|^2 = \left|\left[\mathbf{O}_{d}^{1/2} \mathbf{a}\right]_{k}\right|^2 = \left|\langle {k} | {\alpha} \rangle\right|^2, \quad k = 1, \dots, d.
\end{equation}
Since these weights correspond to the diagonal elements of the density matrix in the Löwdin (orthonormal) basis, they are probabilities by definition, satisfying \(w_k^{\text{L}} (\alpha) \geq 0\) and \(\sum_{k=1}^{d} w_k^{\text{L}} (\alpha) = 1\). They capture the \textit{effective contribution} of each non-orthogonal basis vector to the physical state, interpreted through its orthonormal Löwdin image.  More specifically, Eq.~\eqref{LöwdinWeightsD} shows how the original coefficients \(\left\{a_k\right\}_{k=1}^{d}\), along with the pairwise overlaps \(\langle{c_i} | {c_j}\rangle\) (\(i \neq j\)), determine the distribution of the state \(\ket{\alpha}\) over the Löwdin basis. From another perspective, these weights quantify how strongly the state \(\ket{\alpha}\) aligns with the orthonormal directions \(\ket{k}\), providing a consistent generalization of probabilistic weights to the non-orthogonal setting, while preserving geometric and physical interpretability.

We can extend the preceding analysis from a pure superposition state \(\ket{\alpha}\), given in Eq.~\eqref{Sup-Alpha}, to a general density operator \(\hat{\rho}\) supported on the span of the non-orthogonal basis \(\{\ket{c_k}\}_{k=1}^d\). For a density operator $\hat{\rho}$ represented by the ``coefficient matrix'' $\boldsymbol{\rho}$ in the basis $\{\ket{c_i}\}$:
\begin{equation}
    \hat{\rho} = \sum_{i,j=1}^{d} [\boldsymbol{\rho}]_{ij} \ket{c_i}\bra{c_j},
\end{equation}
the Löwdin-transformed matrix $\boldsymbol{\rho}_{\text{L}}$ represents the statistical weights in the corresponding orthonormal basis. The transformation is strictly defined in terms of the coefficient matrix \(\boldsymbol{\rho}\) as:
\begin{eqnarray}\label{LöwdinWeightsRho}
w_k^{\text{L}}(\hat{\rho}) = \left[\boldsymbol{\rho}_{\text{L}}\right]_{kk} = \left[\frac{\mathbf{O}_{d}^{1/2} \boldsymbol{\rho} \mathbf{O}_{d}^{1/2}}{\mathrm{Tr}\left(\mathbf{O}_{d}^{1/2} \boldsymbol{\rho} \mathbf{O}_{d}^{1/2}\right)}\right]_{kk}.
\end{eqnarray}
It is important to note here that, when expressed in a non-orthonormal basis, the trace of the density operator \(\hat{\rho}\) takes the form
\begin{equation}
\mathrm{Tr}(\hat{\rho}) = \sum_{i,j}[\boldsymbol{\rho}]_{ij}\langle c_j \mid c_i\rangle = \sum_{i,j}[\boldsymbol{\rho}]_{ij}\,[\mathbf{O}_d]_{ji},
\end{equation}
and therefore cannot, in general, be identified with the simple sum of diagonal coefficients \(\sum_i [\boldsymbol{\rho}]_{ii}\). The weights given by Eq.~\eqref{LöwdinWeightsRho} generalize the pure-state case, preserving key properties:
\begin{itemize}
    \item \textbf{Non-negativity}: \(w_k^{\text{L}}(\hat{\rho}) \geq 0\) and \(\sum_k w_k^{\text{L}}(\hat{\rho}) = 1\) (ensured by normalization \(\mathrm{Tr}(\boldsymbol{\rho}_{\text{L}}) = 1\)).\footnote{\label{fn:ChirgwinCoulson} This property fundamentally distinguishes Löwdin weights from other population analyses, notably the Chirgwin-Coulson coefficients. For a general normalized pure state $\ket{\psi} = \sum_{k=1}^d a_k \ket{c_k}$ expanded in a non-orthogonal basis, the Chirgwin-Coulson coefficients are defined as $q_k = \sum_{j=1}^d [\mathbf{O}_d]_{kj}a_k^*a_j$. To illustrate a critical limitation of this approach, consider the specific two-dimensional case with real overlap $s=0.5$ and the normalized state $\ket{\psi} = (0.2\ket{c_1} - \ket{c_2})/\sqrt{0.84}$. A direct calculation yields $q_1 \approx - 0.07$ and $q_2 \approx 1.07$. The emergence of a negative value for $q_1$ demonstrates that these coefficients cannot be interpreted as physical probabilities. In contrast, the Löwdin prescription guarantees a non-negative probability distribution by construction.}
    \item \textbf{Consistency}: For a pure state \(\hat{\rho} = \ket{\alpha}\bra{\alpha}\), \(w_k^{\text{L}}(\hat{\rho}) = w_k^{\text{L}}(\alpha) = \left|\langle {k} | {\alpha} \rangle\right|^2\).
    \item \textbf{Uniqueness}: The Löwdin weights are uniquely determined by the set of initial non-orthogonal basis vectors. Unlike other orthogonalization schemes (e.g., Gram-Schmidt), the LSO approach eliminates arbitrary phase and ordering freedoms, thereby yielding a canonical orthonormal reference frame.
\end{itemize}
The Löwdin weights \(w_k^{\text{L}}(\hat{\rho})\) thus provide a natural probabilistic interpretation for non-orthogonal expansions, reducing to standard Born-rule probabilities \cite{Brumer2006, Zurek2011} when the basis set \(\{\ket{c_k}\}_{k=1}^d\) is orthogonal (i.e., \(\mathbf{O}_d = \mathbf{I}_d\)). In this context, it is worth emphasizing that related quantities such as Chirgwin-Coulson coefficients \cite{Chirgwin50, Torun21} lack this strict non-negativity guarantee, which precludes their interpretation as physical probabilities.

Moreover, the Löwdin weights \cite{LowdinW1953} offer significant advantages in various areas of quantum theory. Their theoretical and practical significance is summarized in four key aspects:
\begin{enumerate}
    \item[(1)] \textit{Extension of Coherence to Non-Orthogonal Frames}: 
    Standard quantum coherence \cite{Baumgratz14} is strictly defined with respect to fixed orthonormal bases \cite{Aberg2006, Shao2015QuCo, Yu2016QuCo, Zhang2016QuCo, Streltsov2017QuCo, Yu2020QuCo, Bischof2021QuCo, Sun2022QuCo, Deng2022QuCo, Ye2023QuCo, Budiyono2023QuCo, Fan2024QuCo, Soulas2025QuCo}. In non-orthogonal settings, however, the expansion coefficients lack a direct probabilistic interpretation due to basis overlaps. Under the broader framework of superposition \cite{Theurer17}, the Löwdin weights resolve this by mapping the state onto the symmetrically orthogonalized frame. This construction recovers a legitimate probability distribution, allowing standard coherence measures to be consistently applied to non-orthogonal systems without violating physical constraints.

    \item[(2)] \textit{Elimination of Basis-Ordering Ambiguity}: 
    A fundamental issue in non-orthogonal expansions is the path-dependence of the reference frame. Conventional methods, such as Gram-Schmidt orthogonalization, are sensitive to the ordering of basis vectors, yielding different weights for different permutations of the same set. The Löwdin process treats all basis vectors symmetrically, removing this ambiguity. It defines a unique orthonormal reference independent of basis labeling, thereby providing a canonical and order-invariant quantification of component contributions \cite{LOWDIN1950}.

    \item[(3)] \textit{Intrinsic Geometric Universality}: 
    A distinct advantage of the Löwdin formulation is its abstraction from the explicit functional form of the basis vectors. The weights are determined solely by the intrinsic geometry encoded in the overlap matrix $\mathbf{O}_d$ and the coefficient vector. Consequently, the Löwdin analysis is physically universal: it remains invariant under any unitary transformation of the basis set that preserves the overlap structure. This ensures that the characterization depends only on the information-theoretic relations between states, rather than their specific physical realization (e.g., atomic orbitals vs. optical modes).

    \item[(4)] \textit{Minimal Geometric Distortion}: 
    Among all possible orthonormalizations, the Löwdin basis is optimally close to the original non-orthogonal set in the least-squares sense (Eq.~\eqref{Least-Square-Sense}). This proximity ensures that the transformation preserves the maximal amount of physical and geometric structure from the original representation. As a result, the Löwdin weights maintain the strongest possible interpretative continuity with the original non-orthogonal expansion, making them ideal for applications requiring minimal distortion, such as variational analysis.
\end{enumerate}


\section{A Löwdin-Weight Perspective on Superposition and Coherence Characterization}
\label{section-LÖwdinWeights-Examples}

The following set of four examples --- comprising two superposition states (one being the golden state, i.e., a maximal superposition state) and two superposition-free states (one being the maximally mixed state) --- illustrates the application of the Löwdin weights in characterizing superposition and in quantifying the associated degrees of superposition and coherence.

We consider a two-dimensional non-orthogonal basis set \(\{\ket{c_1}, \ket{c_2}\}\) with overlap \(s = 1/2\). The associated overlap matrix is given by:
\begin{eqnarray}\label{Example-Gram2D}
\mathbf{O}_{2} = \begin{pmatrix}
1 & s \\
s & 1
\end{pmatrix} = \begin{pmatrix}
1 & 0.5 \\
0.5 & 1
\end{pmatrix}.
\end{eqnarray}
Based on this metric, we analyze specific illustrative examples. In the first example, we explicitly verify the normalization of a density operator to establish the distinction between coefficient summation and the true operator trace. Consider the density operator \(\hat{\rho}\) defined by the expansion:
\begin{equation}\label{Example-Gram2D1st}
\hat{\rho} = \frac{1}{2} |c_1\rangle\langle c_1| + \frac{1}{6} |c_1\rangle\langle c_2| + \frac{1}{6} |c_2\rangle\langle c_1| + \frac{1}{3} |c_2\rangle\langle c_2|.
\end{equation}
It is instructive to distinguish the coefficient matrix \(\boldsymbol{\rho}\), which contains the raw expansion parameters from Eq.~\eqref{Example-Gram2D1st}, from the ``matrix representation'' \([\hat{\rho}]\). The latter is defined by the physical projections \([\hat{\rho}]_{ij} = \langle c_i | \hat{\rho} | c_j \rangle\) and explicitly incorporates the basis overlaps:
\begin{equation}\label{Two-rho-form}
\boldsymbol{\rho} = \begin{pmatrix} 
1/2 & 1/6 \\ 
1/6 & 1/3 
\end{pmatrix}, \quad 
[\hat{\rho}] = \begin{pmatrix} 
3/4 & 5/8 \\ 
5/8 & 5/8 
\end{pmatrix}.
\end{equation}
We demonstrate that \(\mathrm{Tr}(\hat{\rho}) = 1\) using two complementary approaches. First, we compute the trace directly in Hilbert space by applying the cyclic property \(\mathrm{Tr}(\ket{\psi}\bra{\phi}) = \langle\phi|\psi\rangle\):
\begin{align}
\mathrm{Tr}(\hat{\rho}) &= \frac{1}{2}\langle c_1 | c_1 \rangle + \frac{1}{6}\langle c_2 | c_1 \rangle + \frac{1}{6}\langle c_1 | c_2 \rangle + \frac{1}{3}\langle c_2 | c_2 \rangle \nonumber \\
&= \frac{1}{2}(1) + \frac{1}{6}\left(\frac{1}{2}\right) + \frac{1}{6}\left(\frac{1}{2}\right) + \frac{1}{3}(1) = 1.
\end{align}
Second, we employ the matrix formalism using the coefficient matrix \(\boldsymbol{\rho}\) given in Eq.~\eqref{Two-rho-form} together with the overlap matrix \(\mathbf{O}_{2}\) given in Eq.~\eqref{Example-Gram2D}. The operator trace is obtained as the trace of their product:
\begin{equation}
\mathrm{Tr}(\hat{\rho}) = \mathrm{Tr}(\boldsymbol{\rho}\mathbf{O}_{2}) = \mathrm{Tr}\left[
\begin{pmatrix} 
7/12 & 5/12 \\ 
1/3 & 5/12 
\end{pmatrix}
\right] = \frac{7}{12} + \frac{5}{12} = 1.
\end{equation}
Note that while the matrix representation \([\hat{\rho}]\) \eqref{Two-rho-form} has a diagonal sum of \(3/4 + 5/8 = 11/8 \neq 1\), this is expected; in non-orthogonal bases, the simple sum of diagonal values does not correspond to the physical trace.

Following the Löwdin's SO procedure detailed in Section \ref{section-LSymmetric}, the square root of the overlap matrix is calculated using the eigenvalues \(\lambda_{1,2} = 1 \pm s\):
\begin{eqnarray}\label{SquareRootGramD2}
\mathbf{O}_{2}^{1/2} = \begin{pmatrix}
0.966 & 0.259 \\
0.259 & 0.966
\end{pmatrix}.
\end{eqnarray}
Applying the transformation \eqref{LöwdinWeightsRho} to the coefficient matrix \(\boldsymbol{\rho}\) yields the density operator in the orthonormal Löwdin basis:
\begin{eqnarray}\label{Example-Gram2D1stL}
\boldsymbol{\rho}_{\text{L}} = \frac{\mathbf{O}_{2}^{1/2} \boldsymbol{\rho} \mathbf{O}_{2}^{1/2}}{\mathrm{Tr}\left(\mathbf{O}_{2}^{1/2} \boldsymbol{\rho} \mathbf{O}_{2}^{1/2}\right)} = 
\begin{pmatrix}
0.572 & 0.375 \\
0.375 & 0.428
\end{pmatrix}.
\end{eqnarray}
The Löwdin weights are given by the diagonal elements of \(\boldsymbol{\rho}_{\text{L}}\): \(w_1^{\text{L}}(\rho) \approx 0.572\) and \(w_2^{\text{L}}(\rho) \approx 0.428\). These weights sum to unity, forming a valid probability distribution that correctly accounts for the non-orthogonal metric. We examine three further examples before assessing the implications of Eq.~\eqref{Example-Gram2D1stL}.

We now present the second and third examples: two distinct superposition-free states. Recall that a state is superposition-free if it admits the form \(\hat{\rho} = \sum_k p_k \ket{c_k}\bra{c_k}\) with \(p_k \geq 0\) \cite{Theurer17}. For our second example, let the state be defined by the coefficients \(p_1 = 5/6\) and \(p_2 = 1/6\):
\begin{eqnarray}\label{Example-Gram2D2nd}
\hat{\rho} = \frac{5}{6} \ket{c_1}\bra{c_1} + \frac{1}{6} \ket{c_2}\bra{c_2}.
\end{eqnarray}
The corresponding coefficient matrix \(\boldsymbol{\rho}\) and the matrix representation \([\hat{\rho}]\) in the non-orthogonal basis are:
\begin{equation}
\boldsymbol{\rho} = \begin{pmatrix} 5/6 & 0 \\ 0 & 1/6 \end{pmatrix}, \quad 
[\hat{\rho}] = \begin{pmatrix} 7/8 & 1/2 \\ 1/2 & 3/8 \end{pmatrix}.
\end{equation}
Before proceeding, a crucial clarification is necessary regarding these matrices. The state \(\hat{\rho}\) in Eq.~\eqref{Example-Gram2D2nd} is explicitly a classical mixture, and thus superposition-free \cite{Theurer17} with respect to the basis \(\{|c_1\rangle, |c_2\rangle\}\). The nonzero off-diagonal elements in the representation \([\hat{\rho}]\) arise not from quantum coherence, but from the inherent non-orthogonality of the basis states themselves (\(\langle c_1|c_2\rangle \neq 0\)). In a non-orthogonal basis, a ``diagonal'' density operator (representing a classical statistical mixture) generally possesses a non-diagonal matrix representation. Therefore, the off-diagonal terms in \([\hat{\rho}]\) are a direct consequence of the metric geometry, not the quantum state.

By following the same procedure as in the previous example, one verifies that \(\mathrm{Tr}(\hat{\rho}) = 1\). Applying the transformation \eqref{LöwdinWeightsRho} yields:
\begin{eqnarray}\label{Example-Gram2D2ndL}
\boldsymbol{\rho}_{\text{L}} =
\begin{pmatrix}
0.789 & 0.250 \\
0.250 & 0.211
\end{pmatrix}.
\end{eqnarray}
The Löwdin weights \(w_1^{\text{L}}(\hat{\rho}) \approx 0.789\) and \(w_2^{\text{L}}(\hat{\rho}) \approx 0.211\) maintain the unit trace. A crucial observation is that while \(\hat{\rho}\) given in Eq.~\eqref{Example-Gram2D2nd} is superposition-free in the non-orthogonal basis \(\{|c_1\rangle, |c_2\rangle\}\), it exhibits quantum coherence (off-diagonal terms) in the Löwdin basis. This occurs because the Löwdin transformation reveals the coherence implicit in the original representation: each non-orthogonal basis state \(\ket{c_k}\) is itself a coherent superposition when viewed from the orthonormal Löwdin frame. Consequently, even a classical mixture in the non-orthogonal basis manifests as a coherent state in the orthogonal framework.

Our third example is the maximally mixed state, which is also superposition-free in the two-dimensional space defined by \(\{|c_1\rangle, |c_2\rangle\}\). It is given by equal coefficients:
\begin{eqnarray}\label{Example-Gram2D3rd}
\hat{\rho} = \frac{1}{2} \ket{c_1}\bra{c_1} + \frac{1}{2} \ket{c_2}\bra{c_2}.
\end{eqnarray}
The coefficient matrix is \(\boldsymbol{\rho} = \frac{1}{2}\mathbf{I}\), and the matrix representation is \([\hat{\rho}] = \begin{pmatrix} 5/8 & 1/2 \\ 1/2 & 5/8 \end{pmatrix}\). A straightforward derivation yields the Löwdin state:
\begin{eqnarray}\label{Example-Gram2D3rdL}
\boldsymbol{\rho}_{\text{L}} = \frac{1}{2}\mathbf{O}_{2} = 
\begin{pmatrix}
0.50 & 0.25 \\
0.25 & 0.50
\end{pmatrix}.
\end{eqnarray}
The Löwdin weights \(w_1^{\text{L}}(\hat{\rho}) = 0.5\) and \(w_2^{\text{L}}(\hat{\rho}) = 0.5\) remain equal and normalized. Comparing the second and third examples reveals that the off-diagonal components of the density matrices in Eqs.~\eqref{Example-Gram2D2ndL} and \eqref{Example-Gram2D3rdL} are identical (\(0.25\)). This equality arises because, for any diagonal coefficient matrix \(\boldsymbol{\rho}\), the off-diagonal term in the Löwdin basis is proportional to the trace of \(\boldsymbol{\rho}\) (which is unity for both states) scaled by the basis overlap.

For our fourth example, we turn our attention to the maximal superposition state. Based on the overlap \(s=1/2\) and the associated metric specified in Eq.~\eqref{Example-Gram2D}, the maximal superposition state (often referred to as the Golden State \cite{Torun23Low}) is defined as the normalized pure state \(\ket{\Psi_2^-} = \ket{c_1} - \ket{c_2}\). The corresponding density operator defined by the coefficient matrix \(\boldsymbol{\rho}\) is:
\begin{eqnarray}\label{Example-Gram2D4th}
\hat{\rho} = \ket{c_1}\bra{c_1} - \ket{c_1}\bra{c_2} - \ket{c_2}\bra{c_1} + \ket{c_2}\bra{c_2}, \quad 
\boldsymbol{\rho} = \begin{pmatrix}
1 & -1 \\
-1 & 1
\end{pmatrix}.
\end{eqnarray}
We note that the trace is correctly unity: \(\mathrm{Tr}(\hat{\rho}) = \mathrm{Tr}(\boldsymbol{\rho}\mathbf{O}_2) = 1\). The transformation \eqref{LöwdinWeightsRho} then yields the orthonormal representation:
\begin{eqnarray}\label{Example-Gram2D4thL}
\boldsymbol{\rho}_{\text{L}} = 
\begin{pmatrix}
0.5 & -0.5 \\
-0.5 & 0.5
\end{pmatrix}.
\end{eqnarray}
The Löwdin weights \(w_1^{\text{L}}(\hat{\rho}) = 0.5\) and \(w_2^{\text{L}}(\hat{\rho}) = 0.5\) represent a balanced distribution. Guided by these four illustrative examples, our analysis unfolds in two distinct stages: first, a characterization of the superposition states based on the structural properties of the transformed density matrix \(\boldsymbol{\rho}_{\text{L}}\), and second, a quantitative evaluation of their degrees of superposition and coherence. A summary of the four examples examined is provided in Table~\ref{table2}.

\begin{table}[htb!]
    \centering
    \caption{Summary of the four illustrative quantum states in the non-orthogonal basis ($\{\ket{c_i}\}$ with overlap $s = 1/2$) and their properties in the orthonormal Löwdin basis. The table contrasts the state definition with the resulting Löwdin weights ($w^{\text{L}}_k = [\boldsymbol{\rho}_{\text{L}}]_{kk}$) and off-diagonal coherence ($\left|[\boldsymbol{\rho}_{\text{L}}]_{12}\right|$).}
    \label{table2}
    \renewcommand{\arraystretch}{1.3} 
    \setlength{\tabcolsep}{8pt}       
    \begin{tabular}{@{}c l c c c@{}} 
    \toprule
    \textbf{Ex.} & \textbf{State Characterization} & \textbf{Ref.} & \textbf{Löwdin Weights} $(w^{\text{L}}_1, w^{\text{L}}_2)$ & $\left|[\boldsymbol{\rho}_{\text{L}}]_{12}\right|$ \\ 
    \midrule
    \textbf{1} & Superposition State & \eqref{Example-Gram2D1st} & $(0.572, \, 0.428)$ & $0.375$ \\ 
    \textbf{2} & Classical Mixture & \eqref{Example-Gram2D2nd} & $(0.789, \, 0.211)$ & $0.250$ \\ 
    \textbf{3} & Maximally Mixed State & \eqref{Example-Gram2D3rd} & $(0.500, \, 0.500)$ & $0.250$ \\ 
    \textbf{4} & Golden State & \eqref{Example-Gram2D4th} & $(0.500, \, 0.500)$ & $0.500$ \\ 
    \bottomrule
    \end{tabular}
\end{table}

\textbf{(1) Characterization of Superposition States:} The Löwdin transformation provides a privileged framework for characterizing quantum superposition in non-orthogonal systems. A crucial observation emerges from examining the off-diagonal elements of \(\boldsymbol{\rho}_{\text{L}}\) across our examples (summarized in Table~\ref{table2}), which establishes a fundamental detection scale dictated by the specific overlap \(s = 1/2\).

\begin{figure}[htb!]
\centering
\includegraphics[width=12.50cm]{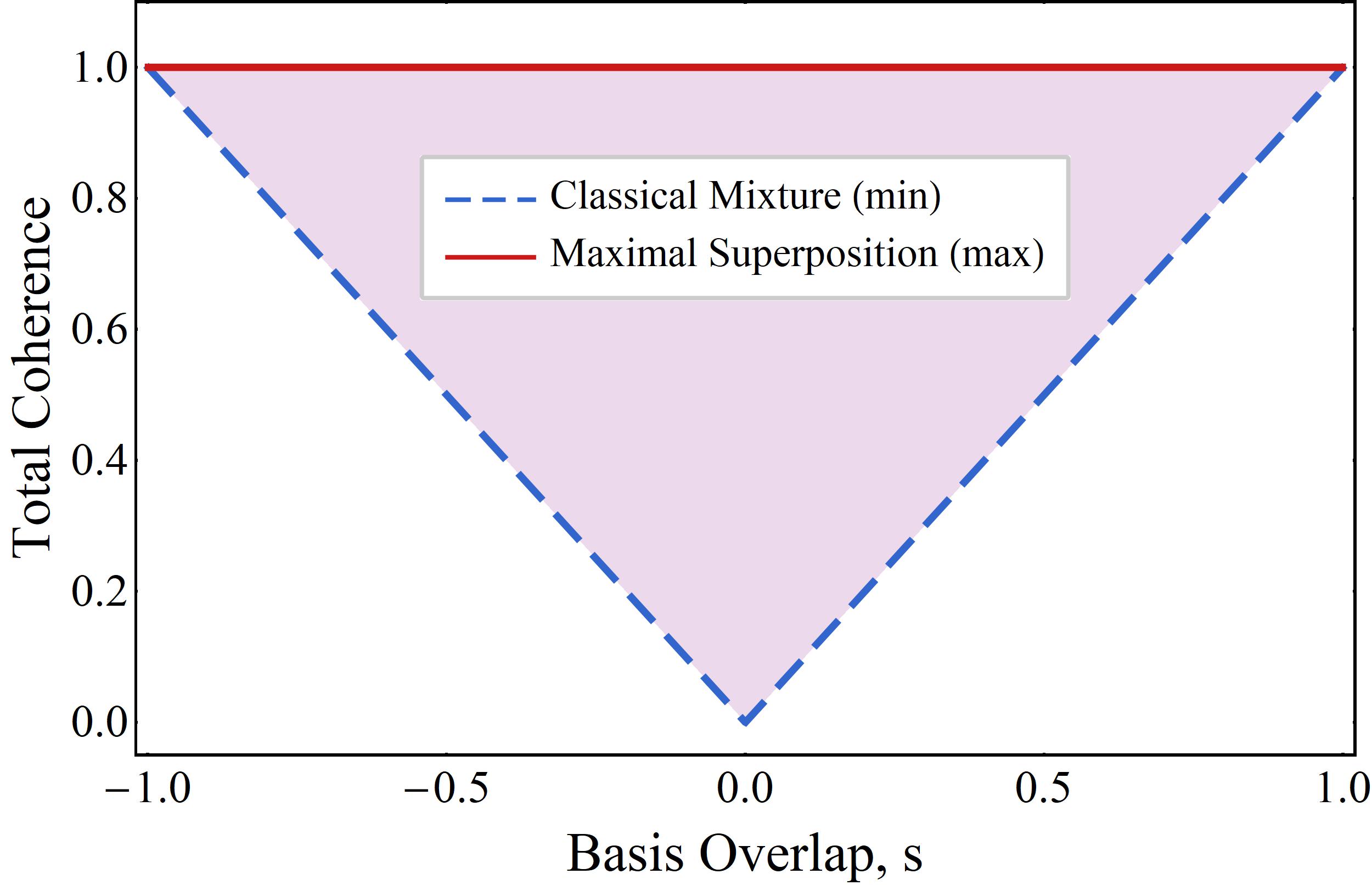}
\caption{Total coherence as a function of the basis overlap parameter $s$ in the Löwdin representation. The shaded region delimits the accessible coherence space, encompassing the full continuum of valid quantum states --- both superposition and superposition-free regimes. The lower bound (dashed blue curve) defines the superposition-free limit (classical mixtures), quantifying the irreducible ``geometric coherence'' imposed solely by the non-orthogonality of the basis. The upper bound (solid red line) marks the saturation limit corresponding to maximal superposition states, where the coherence reaches its theoretical maximum of unity \cite{Torun23Low}. Note that the specific form of the maximal state transitions from the symmetric combination $\ket{c_1} + \ket{c_2}$ for $s < 0$ to the antisymmetric form $\ket{c_1} - \ket{c_2}$ for $s > 0$.}
\label{FIG:TotCoherence}
\end{figure}

\begin{itemize}
    \item \textbf{Superposition-Free Signature (Baseline Coherence):} The two superposition-free states (Examples 2 and 3) yield identical absolute off-diagonal values of \(|[\boldsymbol{\rho}_{\text{L}}]_{12}| = |[\boldsymbol{\rho}_{\text{L}}]_{21}| = 0.250\). This numerical equivalence is not coincidental but reflects a fundamental geometric property. For a fixed overlap \(s\), any normalized classical mixture (i.e., any state with a diagonal coefficient matrix \(\boldsymbol{\rho}\)) transforms into a Löwdin state with off-diagonal magnitude exactly equal to \(\omega = s/2\). Consequently, the value \(\omega = 0.250\) serves as the \emph{baseline coherence} signature. This represents the geometric coherence induced purely by the non-orthogonality of the basis, independent of any quantum superposition in the state itself.

    \item \textbf{Maximal Superposition Bound (Upper Limit):} Conversely, the golden state (Example 4), defined as the maximal superposition state, achieves the theoretical maximum with \(\left|[\boldsymbol{\rho}_{\text{L}}]_{12}\right| = \left|[\boldsymbol{\rho}_{\text{L}}]_{21}\right| = 0.500\). This value, \(\beta = 0.500\), represents the \emph{upper bound} for coherence in this two-dimensional Löwdin basis.

    \item \textbf{Intermediate Superposition States:} The superposition state (Example 1), with its off-diagonal value of \(0.375\), occupies the interval between these critical bounds (\(\omega < 0.375 < \beta\)). This positioning quantitatively verifies its nature as a ``genuine'' quantum superposition state. It exhibits coherence exceeding the geometric baseline (\(0.375 > \omega\)), confirming the presence of constructive quantum interference, yet remains below the limit of the maximal state (\(0.375 < \beta\)).
\end{itemize}
This characterization establishes that the off-diagonal element \(|[\boldsymbol{\rho}_{\text{L}}]_{12}|\) serves as a sensitive indicator of superposition type. The values \(\omega\) and \(\beta\) create a calibrated scale where the position of a state's coherence magnitude directly reveals its superposition properties relative to the non-orthogonal basis \(\{\ket{c_1}, \ket{c_2}\}\). Figure \ref{FIG:TotCoherence} illustrates the coherence magnitude within superposition and superposition-free states, characterized by their respective lower and upper bounds.

\textbf{(2) Quantification of Superposition and Coherence:} The structural characterization naturally leads to quantitative measures for assessing the degree of superposition and coherence.

\begin{figure}[htb!]
\centering
\includegraphics[width=12.50cm]{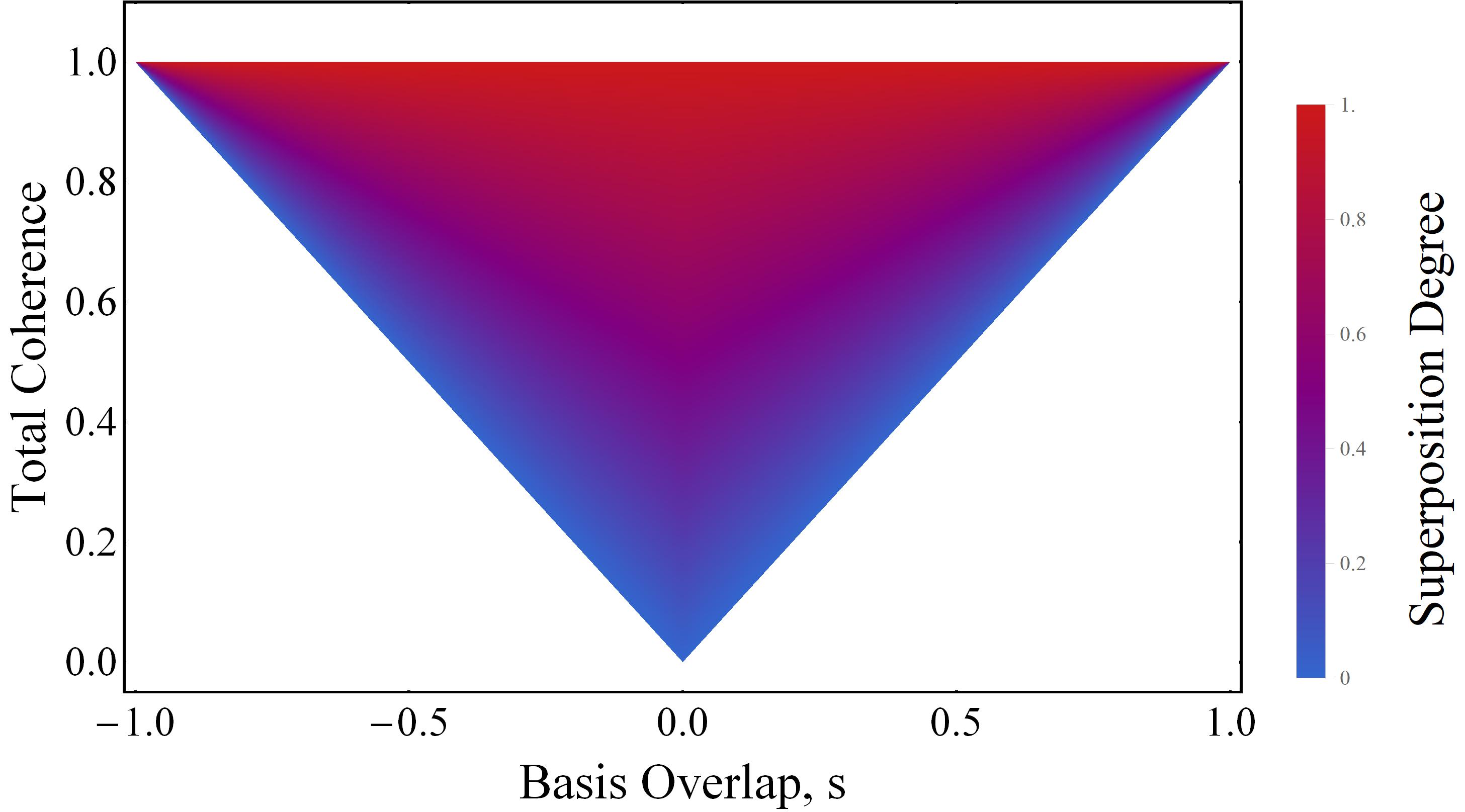}
\caption{Density plot quantifying the superposition degree $\mathcal{S}_{\text{off}}$ (Eq.~\eqref{ASupDegreeMeasure}) within the accessible parameter space of the Löwdin representation. The color gradient maps the normalized proximity to the maximal superposition limit, scaling from superposition-free classical mixtures (blue, $\mathcal{S}_{\text{off}}=0$) to maximal superposition states (red, $\mathcal{S}_{\text{off}}=1$). The accessible region is delimited by the irreducible geometric coherence $|s|$ (lower bound) and the saturation limit of unity (upper bound). Thus, the gradient visualizes the emergence of genuine quantum resources (i.e., superposition) exceeding the baseline coherence inherent to the non-orthogonal basis.}
    \label{FIG:SuperpositionDegree}
\end{figure}

\begin{itemize}
    \item \textbf{Relative Superposition Measure:} The relative position within the established bounds suggests a natural measure for the degree of superposition (see Figure \ref{FIG:SuperpositionDegree}). We define the \emph{relative superposition measure} (for 2D) as:
    \begin{equation}\label{ASupDegreeMeasure}
    \mathcal{S}_{\text{off}}(\hat{\rho}) = \frac{|[\boldsymbol{\rho}_{\text{L}}]_{12}| - \omega}{\beta - \omega},
    \end{equation}
    where \(\omega = 0.25\) represents the superposition-free baseline and \(\beta = 0.5\) the maximal upper bound. This normalized measure yields \(\mathcal{S}_{\text{off}} = 0\) for all superposition-free states and \(\mathcal{S}_{\text{off}} = 1\) for the golden (superposition) state. For Example 1, we compute:
    \begin{equation}
    \mathcal{S}_{\text{off}} \approx \frac{0.375 - 0.250}{0.500 - 0.250} = 0.500.
    \end{equation}
    This result quantitatively confirms that Example 1 represents a state exactly halfway between a classical mixture and maximal superposition in terms of Löwdin coherence.

    \item \textbf{Coherence Quantification via \(l_1\)-norm:} The standard \(l_1\)-norm of coherence \cite{Baumgratz14} provides a complementary measure of the total quantum coherence (see Figure \ref{FIG:TotCoherence}) in the Löwdin basis:
    \begin{equation}
    C_{l_1}(\hat{\rho}) = \sum_{i \neq j} |[\boldsymbol{\rho}_{\text{L}}]_{ij}| = 2|[\boldsymbol{\rho}_{\text{L}}]_{12}|.
    \end{equation}
    The calculated values systematically reflect the superposition hierarchy: Examples 2 \& 3 (superposition-free) yield \(C_{l_1}(\hat{\rho}) = 0.50\); Example 1 yields \(C_{l_1}(\hat{\rho}) = 0.75\); and Example 4 yields \(C_{l_1}(\hat{\rho}) = 1.00\). Crucially, the nonzero coherence (\(C_{l_1}(\hat{\rho}) = 0.50\)) for classical mixtures represents the irreducible coherence inherent in the non-orthogonal basis structure, which becomes explicit only through the Löwdin transformation.
\end{itemize}
This two-tiered quantification demonstrates that while the \(l_1\)-norm measures total coherence in the Löwdin basis, the relative superposition measure \(\mathcal{S}_{\text{off}}\) specifically isolates the additional coherence arising from genuine quantum superposition, filtering out the geometric artifacts of the basis. Together, they provide a comprehensive framework for characterizing and quantifying quantum resources in non-orthogonal systems.

In fact, the preceding examples can be readily generalized to provide a comprehensive analytical foundation. We consider a general density operator \(\hat{\rho}\) in a two-dimensional space spanned by non-orthogonal states, defined by the coefficient matrix \(\boldsymbol{\rho}\):
\begin{eqnarray}\label{Example-Gram2DGenRho}
\hat{\rho} &=& p_1 \ket{c_1}\bra{c_1} + q \ket{c_1}\bra{c_2} + q^* \ket{c_2}\bra{c_1} + p_2 \ket{c_2}\bra{c_2},
\end{eqnarray}
where \(p_1, p_2 \in \mathds{R}_{\geq 0}\) and \(q\) is a complex parameter. The basis overlap is denoted by \(\langle c_1 | c_2 \rangle = s\). The normalization condition \(\mathrm{Tr}(\hat{\rho}) = 1\) imposes the constraint:
\begin{equation}\label{Example-Gram2DGenRho-Normal}
\mathrm{Tr}(\boldsymbol{\rho}\mathbf{O}_2) = p_1 + p_2 + 2s\Re(q) = 1.
\end{equation}
Assuming, without loss of generality, that \(q\) is real-valued, the exact Löwdin transformation \eqref{LöwdinWeightsRho} yields compact analytical expressions for the matrix elements of \(\boldsymbol{\rho}_{\text{L}}\). Let \(\mathcal{P} = p_1 + p_2\) denote the total diagonal population and \(\Delta = p_1 - p_2\) the population difference. The Löwdin weights are derived as:
\begin{eqnarray}
w_1^{\text{L}}(\hat{\rho}) = \frac{1}{2}\left( \mathcal{P} + \Delta\sqrt{1-s^2} \right) + qs,
\end{eqnarray}
\begin{eqnarray}
w_2^{\text{L}}(\hat{\rho}) = \frac{1}{2}\left( \mathcal{P} - \Delta\sqrt{1-s^2} \right) + qs.
\end{eqnarray}
Summing these yields \(w_1^{\text{L}} + w_2^{\text{L}} = \mathcal{P} + 2qs = 1\), confirming consistent normalization (Eq.~\eqref{Example-Gram2DGenRho-Normal}). The most profound insight, however, lies in the structure of the off-diagonal coherence term:
\begin{eqnarray}\label{eq:decomposition_exact}
\left[\boldsymbol{\rho}_{\text{L}}\right]_{12} = \frac{s \mathcal{P}}{2} + q.
\end{eqnarray}
For this two-dimensional system, this result provides a rigorous physical decomposition of coherence in the orthonormal frame. The term \([\boldsymbol{\rho}_{\text{L}}]_{12}\) cleanly separates into two distinct contributions:
\begin{equation}\label{eq:decomposition}
[\boldsymbol{\rho}_{\text{L}}]_{12} = \underbrace{\frac{s \mathcal{P}}{2}}_{\text{Geometry-Induced Coherence}} + \underbrace{q}_{\text{Intrinsic Quantum Superposition}}.
\end{equation}
The first term depends only on the classical populations and the basis overlap \(s\). It is non-zero even for superposition-free states (where \(q=0\)). For a normalized classical mixture (where \(\mathcal{P}=1\)), this term fixes the baseline coherence exactly at \(\omega(s) = s/2\), matching our numerical observations in Table~\ref{table2}. The second term is exactly the parameter \(q\), which quantifies the genuine quantum superposition in the coefficient matrix.

Figure \ref{FIG:SuperpositionDegree} illustrates the superposition degree (Eq.~\eqref{ASupDegreeMeasure}) for valid quantum states, delineated by the coherence boundaries of superposition-free states and maximal superposition states. 
This analysis culminates in a powerful synthesis regarding the characterization and quantification of quantum states in non-orthogonal metric spaces. Fundamentally, the Löwdin transformation rectifies the probabilistic ambiguity of the original coefficient set, yielding weights $\{w_k^\text{L}(\hat{\rho})\}$ that serve as a consistent probability distribution respecting the underlying Hilbert space metric. Crucially, this framework allows for an operational separation of total coherence into geometric and quantum components. We rigorously decompose the off-diagonal terms in the Löwdin basis as:
\begin{equation}
    \left|[\boldsymbol{\rho}_{\text{L}}]_{12}\right| = \left| \underbrace{\omega(s, \mathcal{P})}_{\text{Geometric Artifact}} + \underbrace{\delta(\hat{\rho})}_{\text{Genuine Resource}} \right|,
\end{equation}
where $\omega(s, \mathcal{P}) = s\mathcal{P}/2$ represents the inevitable geometric coherence floor dictated solely by the basis overlap $s$ and total population $\mathcal{P}$. In contrast, $\delta(\hat{\rho}) = q$ isolates the excess coherence arising from intrinsic quantum superposition. Consequently, a non-zero value of $\delta(\hat{\rho})$ emerges as the definitive signature of a resourceful (superposition) state, distinct from the background correlations imposed by the non-orthogonal geometry.

This structural decomposition has profound implications for resource theories of superposition \cite{Baumgratz14, Theurer17}. By subtracting the geometric baseline $\omega$, the Löwdin framework enables the precise isolation and quantification of state-dependent quantum resources, filtering out the fixed representational artifacts of the basis. The consistency of this approach is further validated in the limiting case of an orthonormal basis ($s \to 0$). In this limit, the geometric baseline vanishes ($\omega \to 0$), and the expressions recover the standard quantum mechanical results: the weights converge to the raw coefficients ($w_k^{\text{L}} \to p_k$) and the coherence term simplifies to the raw off-diagonal element ($[\boldsymbol{\rho}_{\text{L}}]_{12} \to q$). This asymptotic behavior confirms that the Löwdin procedure constitutes the correct and consistent generalization of quantum state representation for non-orthogonal systems.


\section{Entropic and Geometric Quantification of Localization}
\label{section-EntropicandGeometric}

Having established the Löwdin representation as the framework for separating geometric artifacts from genuine quantum coherence\footnote{Here, ``genuine'' denotes quantum resources arising from intrinsic superposition, distinct from the baseline coherence induced by non-orthogonality. As shown in Section \ref{section-LÖwdinWeights-Examples}, even classical mixtures exhibit a geometric coherence floor in the Löwdin representation; genuine coherence is defined as the excess magnitude strictly exceeding this irreducible limit.}, we now turn our attention to the diagonal elements of the transformed density matrix \eqref{LöwdinWeightsRho}, \(\mathbf{w} = \operatorname{diag}(\boldsymbol{\rho}_{\text{L}})\). These elements, the Löwdin weights, constitute a valid probability distribution that respects the metric\footnote{This implies that the weights explicitly incorporate the overlap matrix \(\mathbf{O}_{d}\) (the metric tensor of the basis). Unlike raw coefficient populations which disregard cross-terms in non-orthogonal bases, the Löwdin weights ensure that the normalization condition \(\sum_k w_k^{\text{L}} = 1\) holds exactly, providing a physically consistent probability measure.} of the underlying Hilbert space. In this section, we utilize this distribution to quantify the degree of localization and information content of superposition states, employing measures that complement the coherence metrics derived in Section \ref{section-LÖwdinWeights-Examples}.

\subsection{Information-Theoretic Quantification: Entropy as a Measure of Localization}
The distribution \(\mathbf{w} = \{w_k^{\text{L}}\}\) allows for a direct information-theoretic quantification of the state's delocalization across the orthonormalized basis. We define the \emph{Löwdin entropy} as the Shannon entropy of these weights:
\begin{equation}\label{ShannonEntropyLW}
H(\mathbf{w}) = -\sum_{k=1}^d w_k^{\text{L}} \log_2 w_k^{\text{L}}.
\end{equation}
It is crucial to clarify the physical meaning of this quantity \eqref{ShannonEntropyLW}. While the off-diagonal elements of the transformed matrix \(\boldsymbol{\rho}_{\text{L}}\) capture phase relations (coherence), the entropy \(H(\mathbf{w})\) depends solely on the diagonal populations. Therefore, physically, \(H(\mathbf{w})\) measures the localization of the state within the Löwdin basis. The relationship between this localization and the underlying quantum resources depends fundamentally on the purity of the state:

\textbf{(1) For Pure States:}
In the case of a pure superposition state \(\hat{\rho} = \ket{\psi}\bra{\psi}\), the entropic delocalization becomes formally equivalent to the resource-theoretic content of superposition. Since a pure state cannot be delocalized across a basis without exhibiting superposition, \(H(\mathbf{w})\) serves as a direct quantifier of the superposition resource. Mathematically, we assert the following identity linking the superposition of \(\hat{\rho}\) to the coherence of its Löwdin representation \(\boldsymbol{\rho}_{\text{L}}\):
\begin{equation}\label{PureStateIdentity}
\mathcal{S}_{\text{rel}}(\hat{\rho}) \equiv C_{\text{rel}}(\boldsymbol{\rho}_{\text{L}}) = H(\mathbf{w}).
\end{equation}
The justification for this identity proceeds as follows: The relative entropy of coherence for the transformed state is defined as \(C_{\text{rel}}(\boldsymbol{\rho}_{\text{L}}) = S([\boldsymbol{\rho}_{\text{L}}]_{\text{diag}}) - S(\boldsymbol{\rho}_{\text{L}})\) \cite{Baumgratz14}.
First, the term \([\boldsymbol{\rho}_{\text{L}}]_{\text{diag}}\) is the diagonal matrix formed by the weights \(\mathbf{w}\); its von Neumann entropy is exactly the Shannon entropy, \(S([\boldsymbol{\rho}_{\text{L}}]_{\text{diag}}) = H(\mathbf{w})\).
Second, since the transformation given by Eq.~\eqref{LöwdinWeightsD} is a unitary similarity transformation of the operator representation, it preserves the purity and spectrum of the density operator. Thus, for a pure state, \(S(\boldsymbol{\rho}_{\text{L}}) = S(\hat{\rho}) = 0\). Substituting these into the definition yields \(C_{\text{rel}}(\boldsymbol{\rho}_{\text{L}}) = H(\mathbf{w})\).
Consequently, for pure states, the Löwdin entropy provides an exact, basis-invariant measure of the superposition inherent in \(\hat{\rho}\).

\textbf{(2) For Mixed States:}
For general mixed states, the entropy \(H(\mathbf{w})\) quantifies the \emph{total} delocalization. This quantity is a composite of two distinct physical sources: the quantum coherence (representing superposition) and classical statistical mixing. The exact relationship is governed by the decomposition:
\begin{equation}\label{EntropyDecomp}
H(\mathbf{w}) = \underbrace{C_{\text{rel}}(\boldsymbol{\rho}_{\text{L}})}_{\substack{\text{Coherence of } \boldsymbol{\rho}_{\text{L}} \\ (\text{Superposition} + \text{Geometry})}} + \underbrace{S(\hat{\rho})}_{\text{Classical Mixing}}.
\end{equation}
Here, \(S(\hat{\rho})\) is the basis-independent von Neumann entropy quantifying impurity. The term \(C_{\text{rel}}(\boldsymbol{\rho}_{\text{L}})\) represents the total coherence in the Löwdin frame, which, as discussed in Section \ref{section-LÖwdinWeights-Examples}, includes both the genuine resource \(\delta(\hat{\rho})\) (i.e., superposition) and the geometric baseline \(\omega\).
Therefore, \(H(\mathbf{w})\) serves as an upper bound to the quantum resource (\(H(\mathbf{w}) \geq C_{\text{rel}}(\boldsymbol{\rho}_{\text{L}})\)).
A high Löwdin entropy indicates that the state is broadly distributed over the basis. However, isolating the quantum contribution requires subtracting the classical entropy \(S(\hat{\rho})\).
Crucially, by utilizing Löwdin weights \(w_k^{\text{L}}\) (which explicitly incorporate the metric \(\mathbf{O}_{d}\)), \(H(\mathbf{w})\) provides a ``geometry-corrected'' measure of localization, strictly avoiding the probabilistic ambiguities inherent in raw coefficient analysis.

\subsection{Geometric Localization: Participation Ratio}
Complementing the entropic view, the effective dimensionality of the state within the Hilbert space can be gauged using the Participation Ratio (PR) \cite{Borgonovi2019PRatio}. Defined via the second moment of the Löwdin distribution, the PR is given by:
\begin{equation}\label{PRdefinition}
\text{PR}(\mathbf{w}) = \frac{1}{\sum_{k=1}^d \left(w_k^{\text{L}}\right)^2}.
\end{equation}
The PR estimates the effective number of Löwdin modes participating in the superposition. It ranges continuously from \(\text{PR}=1\) (perfect localization on a single mode) to \(\text{PR}=d\) (maximal delocalization). Its reciprocal, the Inverse Participation Ratio (IPR) \cite{Murphy2011IPR, Liu2025IPR}, serves as a direct metric for localization strength. Both measures inherit the probabilistic validity of the Löwdin framework, offering a geometric perspective on state delocalization that is strictly normalized.

\subsection{Illustrative Analysis: Interplay of Geometry and Asymmetry}
To elucidate the physical interplay between basis non-orthogonality and entropic localization, we analyze a normalized pure state \(\ket{\beta}\) in the two-dimensional system ($s = \langle c_1 | c_2 \rangle$) introduced in Eq.~\eqref{Example-Gram2D}. We define the state with a tunable asymmetry parameter \(\gamma \in \mathds{R}\):
\begin{eqnarray}\label{Eq:BetaDef}
\ket{\beta} = \mathcal{N} \big( \ket{c_1} + \gamma \ket{c_2} \big),
\end{eqnarray}
where \(\mathcal{N} = (1 + \gamma^2 + 2 \gamma s)^{-1/2}\) ensures normalization. The Löwdin weights for this state are derived analytically as:
\begin{equation}\label{Eq:AnalyticWeights}
w_{1,2}^{\text{L}}(\beta) = \frac{1}{2} \pm \frac{1-\gamma^2}{2(1+\gamma^2+2\gamma s)}\sqrt{1-s^2}.
\end{equation}
This expression reveals a competitive mechanism between the state's intrinsic coefficient asymmetry (\(\gamma\)) and the basis overlap (\(s\)). We examine three physical regimes to illustrate this behavior:

\begin{itemize}
    \item \textbf{The Orthogonal Limit (\(s \to 0\)):}
    In the absence of overlap, the weights recover the standard Born probabilities: \(w_1 \to 1/(1+\gamma^2)\) and \(w_2 \to \gamma^2/(1+\gamma^2)\). Here, the entropic delocalization is determined solely by the coefficient ratio \(\gamma\).

    \item \textbf{Weak Overlap Regime (\(s = 0.1, \gamma = 0.6\)):}
    With a small overlap of \(s = 0.1\), the weights are \(w_1^{\text{L}} \approx 0.715\) and \(w_2^{\text{L}} \approx 0.285\), yielding an entropy of \(H(\mathbf{w}) \approx 0.862\) bits. Since the basis vectors remain largely distinguishable, the asymmetry of the coefficient \(\gamma\) dominates, resulting in lower entropy (higher localization).

    \item \textbf{Moderate Overlap Regime (\(s = 0.4, \gamma = 0.6\)):}
    Increasing the overlap to \(s = 0.4\) (while maintaining fixed \(\gamma\)) shifts the weights to \(w_1^{\text{L}} \approx 0.66\) and \(w_2^{\text{L}} \approx 0.34\). Consequently, the entropy \emph{increases} to \(H(\mathbf{w}) \approx 0.925\) bits. 
\end{itemize}

Comparing the weak and moderate regimes reveals a profound feature of Löwdin analysis: \emph{increasing non-orthogonality tends to mask coefficient asymmetry.} As \(s\) increases, the physical distinguishability of states \(\ket{c_1}\) and \(\ket{c_2}\) diminishes. The symmetric orthogonalization procedure reflects this by ``smearing'' the probability distribution, effectively pushing the weights closer to the equiprobable limit ($0.5/0.5$) to compensate for the spatial overlap.

Thus, localization in non-orthogonal systems is a dual function of state asymmetry (\(\gamma\)) and basis geometry (\(s\)). The Löwdin entropy \eqref{ShannonEntropyLW} correctly captures this duality, treating the overlap not as a perturbation, but as a fundamental geometric constraint on the information capacity of the basis. While the transformation becomes singular in the limit of linear dependence (\(s \to 1\)), for any physically valid basis (\(s < 1\)), this framework provides a continuous, rigorous quantification of how the state occupies the accessible Hilbert space volume.


\section{Conclusion}\label{section-Conc}

In this work, we demonstrated that Löwdin’s SO approach surpasses both its canonical variant and the Gram-Schmidt procedure in analyzing superpositions of orthogonal and non-orthogonal basis states. Unlike sequential methods such as Gram-Schmidt, which introduce artificial asymmetry through basis ordering, LSO preserves the global symmetry and geometric structure of the original basis. This establishes a robust, order-independent correspondence between non-orthogonal and orthonormal representations \cite{Baumgratz14, Theurer17, Senyasa22} (see particularly Ref.~\cite{Torun23Low}), making it uniquely suitable for physical systems where basis states are physically equivalent.

A pivotal outcome of this study is the formalization of Löwdin weights \cite{LowdinW1953} as a rigorous probability distribution.
We explicitly showed that, unlike Chirgwin-Coulson coefficients which can yield unphysical negative values, Löwdin weights are strictly non-negative, satisfying the axioms required for information-theoretic analysis. Building on this probabilistic foundation, we derived a structural decomposition of quantum coherence in the transformed basis. Our analysis revealed that the total coherence comprises two distinct components: an irreducible ``geometric coherence floor'' determined solely by the metric non-orthogonality, and a ``genuine'' component arising from intrinsic quantum superposition \cite{Pusuluk24}. This decomposition enabled the definition of the \emph{relative superposition measure} ($\mathcal{S}_{\text{off}}$), which filters out basis-induced artifacts to quantify the state's true resourcefulness.

Furthermore, we clarified the distinct physical roles of the diagonal and off-diagonal elements in the Löwdin representation.
While off-diagonal terms quantify phase coherence, we demonstrated that the Shannon entropy of the diagonal Löwdin weights provides a measure of \emph{localization} within the geometrically optimized basis \cite{Cohen2016Loc, Styliaris2019Loc}.
Through explicit examples in two- and three-dimensional systems, we validated that this framework consistently characterizes the continuum from classical mixtures to maximal superposition states.

Overall, the LSO method combined with Löwdin weights provides a physically faithful and geometrically transparent toolset for resource-theoretic analyses \cite{Baumgratz14, Theurer17, Streltsov2017QuCo}. By resolving the ambiguities of basis ordering and metric artifacts, this approach offers a clearer interpretation of interference phenomena, such as those observed in photon-number experiments, where the physical results must be independent of the mathematical order of basis orthogonalization.

Future work can extend these techniques to higher-dimensional systems, explore applications in quantum information protocols, and further clarify the role of orthogonalization in QRTs \cite{Chitambar19, Regula19, Kuroiwa20, Torun23Review, Gour24}. Promising directions include utilizing the non-negativity of Löwdin weights to study the distillation of superposition states \cite{Torun21}; exploring alternative descriptions of superposition such as the biorthogonal resource theory perspective \cite{Pusuluk24} to deepen the understanding of coherence in complex systems; and applying these techniques to continuous-variable systems to fully realize the potential of the Löwdin approach in practical quantum metrology and computing tasks.



\end{document}